\documentclass[%
twocolumn,
 amsmath,amssymb,
 aps,
 pre,
superscriptaddress
]{revtex4-2}
\usepackage{color}
\usepackage{graphicx}
\usepackage{dcolumn}
\usepackage{enumerate}
\usepackage{bm}
\usepackage{hyperref}

\usepackage[colorinlistoftodos]{todonotes}

\def\widthprop{27.5mm}
\def\widthvertex{22.5mm}

\newcommand{\SA}{\mathcal{S}}
\newcommand{\DA}{\mathcal{D}}
\newcommand{\WA}{\mathcal{W}}
\newcommand{\eps}{\varepsilon}

\newcommand{\mx}{{\bm x}}
\newcommand{\mk}{{\bm k}}

\newcommand{\mpp}{{\bm p}}
\newcommand{\boldnabla}{{\bm \nabla}}

\newcommand{\tpsi}{ \tilde{\psi}}

\def\dRM{\mathrm{d}}
\def\eRM{\mathrm{e}}
\usepackage{longtable}

\allowdisplaybreaks 

\begin{document}


\title{Field-theoretic Analysis of Directed Percolation: Three-loop Approximation}

\author{Loran Ts. Adzhemyan}
\affiliation{Sankt Petersburg State University, 7/9 Universitetskaya nab., St. Petersburg, 199034, Russian Federation}
\affiliation{Bogolyubov Laboratory of Theoretical Physics, Joint Institute for Nuclear Research, 141980 Dubna, Russian Federation}

\author{Michal Hnati\v{c}}%
\email{hnatic@saske.sk}
\affiliation{Bogolyubov Laboratory of Theoretical Physics, Joint Institute for Nuclear Research, 141980 Dubna, Russian Federation}
\affiliation{Institute of Experimental Physics, Slovak Academy of Sciences, Watsonova 47, 040 14 Ko\v{s}ice, Slovakia}
\affiliation{Faculty of Science, \v{S}af\'{a}rik University, Moyzesova 16, 040 01 Ko\v{s}ice, Slovakia}

\author{Ella V. Ivanova}
\affiliation{ 
    Otto H. York Department of Chemical and Materials Engineering, New Jersey Institute of Technology, 323 Dr. Martin Luther King Jr. Blvd., Newark, New Jersey 07102, USA
}

\author{Mikhail V. Kompaniets}
\email{m.kompaniets@spbu.ru}
\affiliation{Sankt Petersburg State University, 7/9 Universitetskaya nab., St. Petersburg, 199034, Russian Federation}%
\affiliation{Bogolyubov Laboratory of Theoretical Physics, Joint Institute for Nuclear Research, 141980 Dubna, Russian Federation}

\author{Tom\v{a}\v{s} Lu\v{c}ivjansk\'{y}}
\email{tomas.lucivjansky@upjs.sk}
\affiliation{Faculty of Science, \v{S}af\'{a}rik University, Moyzesova 16, 040 01 Ko\v{s}ice, Slovakia}

\author{Luk\'{a}\v{s} Mi\v{z}i\v{s}in}
\email{mizisin@theor.jinr.ru}
\affiliation{Bogolyubov Laboratory of Theoretical Physics, Joint Institute for Nuclear Research, 141980 Dubna, Russian Federation}

\date{\today}

\begin{abstract}
The directed bond percolation is a paradigmatic model in non-equilibrium statistical physics.
 It captures essential physical information on the nature of continuous phase transition between active and absorbing states.
  In this paper, we study this model by means of the field-theoretic formulation with a subsequent renormalization
  group analysis. We calculate all critical exponents needed for the quantitative description of the corresponding
  universality class to the third order in perturbation theory. Using dimensional regularization with
   minimal subtraction scheme, we carry out perturbative calculations in a formally small parameter $\eps$, where $\eps = 4-d$ is a
   deviation from the upper critical dimension $d_c=4$. 
 We use a non-trivial combination of analytical and numerical tools in order to determine ultraviolet divergent parts of Feynman diagrams. 
\end{abstract}

\maketitle


\section{\label{sec:level1} Introduction}

Non-equilibrium processes are prevalent in nature and the majority of observed phenomena are being in 
 some form of non-equilibrium state~\cite{marro_dickman1999,krapivsky_book2010}. 
 Famous examples encompass turbulent flows~\cite{davidson}, pattern formations~\cite{hohenberg1993}, Earth's atmosphere~\cite{marston2012}, and living organisms~\cite{wang2019}.
 A plethora of other examples can be found not only in the realm of physics and biology but also in chemistry, economy, sociology, climatology, and other research areas. The pervasiveness of the non-equilibrium systems raises the importance of their understanding, which
  is of utmost importance both for theoretical and practical applications. The last decades have witnessed  
 rapid progress in the fundamental understanding of non-equilibrium physics.

Non-equilibrium systems are, from their very nature, dynamical and differ profoundly from systems that can be described 
as near-equilibrium models.
Famous members of the latter group are all models of critical dynamics~\cite{Hohenberg1977,Tauber2014}, which share
 a common property known as fluctuation-dissipation relation~\cite{Vasilev2004}.
 In genuine non-equilibrium systems, this relation is violated and the system is called far from equilibrium.
 Moreover, in order for a system to be in a non-equilibrium state, some mechanisms are necessary, which cause
 continuous doping of energy in and out of a system. This can be generated either by  external means or by 
 internal means as well.

From a whole set of possible non-equilibrium problems, an important collection is formed by growth models, which
find a lot of applications in population dynamics, the creation of fractal structures, etc.
 From a theoretical point of view, such models can be described as
stochastic systems in which the microscopic degrees of freedom evolve according to some probabilistic rules. The collective
 behavior of many microscopic entities usually allows us to employ continuum approximation in which certain
 non-universal properties, such as the type of a lattice structure, are absent.
 
Remarkably, despite the notorious difficulty of non-equilibrium models, they might simplify under specific circumstances and can be analytically tractable.
 We anticipate such simplification to occur when a dynamical system undergoes a continuous phase transition akin to equilibrium second-order phase transitions.
 At criticality, underlying degrees of freedom behave collectively over many spatio-temporal scales, which gives rise to self-similar scaling
 behavior. A particular hallmark is a large correlation length with respect to both time and spatial directions.
 In the critical region, it is then permissible~\cite{Zinn2002,Zinn2007} to approximate a system by means of the continuous or mesoscopic approach.
 Averaging small volumes whose diameter is much smaller than the correlation length leads to an effective description in terms of field
 variables. It is well-known~\cite{Zinn2007,Kardar2007} that this enables us to use powerful methods of quantum (statistical) field theory.
Further, many concepts and methods from equilibrium critical models can be taken over. One of the most important is 
 the concept of universality class. According to it, systems can be categorized into different classes, whose members share the same
 macroscopic behavior. Completely different systems from a microscopic point of view
 could display the same critical behavior, which is determined to a great extent
  by some common gross properties such as space dimension, symmetry, nature of order parameter, etc. All systems in the same universality class are quantitatively
   described by the same set of critical exponents. Therefore, to analyze their critical behavior it is advantageous to choose the simplest possible
  member of a given class.
 
Of particular importance is a group of dynamical models, whose phase space allows the existence of so-called absorbing states.
 These are such configurations that, once entered by a system, they cannot be left. 
 The directed bond percolation (DP) process is probably the most common and famous paradigmatic model of fractal growth. Initially, it was
 developed~\cite{Broadbent1957} to model a spreading of fluid through an irregular porous medium in the 
presence of some external force, e.g. gravity. In biological terms, DP can be interpreted as a 
 simple model for the epidemic process \cite{Murray2002} in which there is no immunization
 of infected individuals. In the evolution of infectious diseases, sick individuals can spread infection to neighbors. The infected individuals are
  allowed to recover, but later the individuals can be sick again. The simple model exhibits a critical region between
  the state without the disease (the absorbing state) and the 
  epidemic (the active state).
Later on, several other applications of DP have been found~ \cite{Hinrichsen2000} in research areas as
 diverse as high energy physics~\cite{Moshe1978,Cardy80}, population dynamics~\cite{Tauber2012}, 
 reaction-diffusion problems \cite{Odor2004}, contact process \cite{Liggett2012}, forest fires models \cite{Albano1995}, a transition from laminar to turbulent flow 
   \cite{Pomeau1986, Lemoult2016, Sano2016}, etc.  Numerous other examples can be
    found in different models of physics, chemistry, biology, and even ecology~\cite{Sahini2003}. 
   
Major attributes of the DP universality class are succinctly summarized by Janssen and Grassberger 
conjecture~\cite{Janssen1981, Grassberger1982}, according to which a system  belongs to the DP universality class if four conditions are met:
     (i) the existence of the unique absorbing state, (ii) the positive
     one-component order parameter, (iii) short-range interactions, (iv) no additional specific properties like quenched disorder, or 
     symmetries. Although this DP hypothesis has not been proven  rigorously yet, strong support exists in its favor~\cite{Hinrichsen2000, Lubeck2005, Henkel2008}. 
Systems fulfilling these conditions are expected to exhibit the same critical behavior. Similarly to
  equilibrium critical phenomena, non-equilibrium phase transition can be categorized into a different universality class 
  \cite{Henkel2008, Tauber2014}, where systems within a given class display the same critical behavior.
  It might be claimed that the DP universality class plays a similar role in
   non-equilibrium dynamics as the Ising universal class in equilibrium systems \cite{Hinrichsen2000}. 

Notwithstanding the fact that the DP universality class is quite robust, critical exponents were measured only in
 a few experiments during the past two decades. The DP phase transition was experimentally studied in turbulent liquid crystals 
 \cite{Takeuchi2007, Takeuchi2009}, the transition from laminar to turbulent flow in channel flow \cite{Goldenfeld2011,Sano2016}, and in Couette 
 flow \cite{ Lemoult2016}. Direct experimental verification is surprisingly low, especially as various possible experiments have been suggested in the past. Clearly, the experimental investigation of DP constitutes a formidable task for  non-equilibrium physics.
   
   Despite a lot of effort that has been put into a theoretical
    analysis of DP, even its simplest formulation in $(1+1)$-dimensions remains exactly unsolvable~\cite{Henkel2008}.
Typical theoretical approaches include numerical or approximate techniques
such as Monte-Carlo, simulations, series expansions, 
 diagonalization techniques, mapping on quantum spin chains, and others.

In this paper, we adopt a field-theoretic approach to study the 
universal properties of the directed percolation process. Though far from being the decisive approach to
a problem, this method offers a great insight into the origin of universality and validation of scaling relations. Moreover, what is
most relevant for this work is that it provides us with a calculation framework in which universal quantities can be calculated systematically. 
 Our starting point is a mesoscopic formulation of DP, which further facilitates
 the use of a very powerful framework of the quantum field theory~\cite{Zinn2007,ChaikinLubensky}. As observed by K.~Wilson critical systems, or more generally stochastic dynamical systems,
  can be interpreted as quantum field models in imaginary time, which
  facilitates the application of sophisticated methods from the quantum field theory.
  
First, we reformulate DP into a form of functional integral. Second, we calculate in a perturbative fashion relevant Green functions
in the form of Feynman diagrams. Third, divergent diagrams are  treated with
 a perturbative renormalization group (RG) in order to
gain information about large-scale behavior. RG not only furnishes a conceptual formalism but also
   provides a powerful and versatile mechanism for computing universal quantities as  critical exponents, amplitudes, and even perturbative calculation of scaling functions~\cite{Vasilev2004}. An important parameter in the RG procedure is  the upper 
  critical dimension $d_c$. It turns out that above $d_c$, the mean-field theory, which neglects fluctuations of the order parameter, predicts 
  correct values for the critical exponents~\cite{Amit,Vasilev2004,Zinn2007}. 
   On the other hand, below $d_c$, fluctuations dominate behavior of the critical system and the mean-field approximation is not appropriate. More sophisticated approaches are called for.
 Precisely at the critical dimension $d_c$, RG theory predicts mean-field results with logarithmic corrections~\cite{Vasilev2004}.
    Up to now, the critical exponents have not been solved exactly below $d_c$, not even at  spatial dimension $d=1$. Traditionally, computer simulations are performed for the evaluation of critical exponents of the DP universality class, for instance, Monte Carlo 
     simulation \cite{Henkel2008}. However, in past decades other lattice models were introduced that are members of the DP universality 
     class e.g. the Domany-Kinzel automaton or contact process \cite{Vojta2012}. Also, many numerical methods were developed as $(1+1)$-dimensional
      the low-density series expansion \cite{Jensen1996}, the density matrix renormalization group method \cite{Carlon1999},
%
%
non-perturbative renormalization group \cite{canet2004},
%
%
      and so on. 
      The precision of critical exponents increases in lower space dimensions. However, in higher spatial dimensions below $d_c$, the error bars of computer
       simulations' results did not decrease substantially for many years of active research \cite{Jensen1992, Henkel2008, Tauber2014}.

Within the field-theoretic renormalization group,  a special role is played  by  dimensional regularization combined with 
$\eps$-expansion~\cite{Zinn2002,Vasilev2004}.
Using the former method, we can regularize divergent Feynman diagrams, whereas the latter is a convenient way in which physical expressions are calculated~\cite{Zinn2002,Vasilev2004}.
Formally, a small $\eps$-expansion parameter is given by a difference $d_c-d$ from  the upper critical dimension (as we will see later on (Sec.~\ref{subsec:can_dim})
$d_c=4$ for DP process). The main quantitative results of RG are  asymptotic series for critical exponents, and these have to be properly 
 summed in order to get precise numerical estimations~\cite{BenderBook}. 
 Currently, analytical predictions for the critical exponents of DP universality class are known up
 to the second-order of the perturbation theory \cite{Janssen1981, Janssen2001,Janssen2005,Tauber2014}. 
The main aim of this paper is to substantially extend existing perturbative results up to
 the third order of perturbation theory in $\eps$. To the best of our
 knowledge, this is the first time results with such precision are presented. 
Let us note that multi-loop calculations usually represent a formidable task. Typically, perturbative calculations
are feasible only for a small number of loops. The majority of calculations for non-equilibrium models
are limited only to the two-loop approximation. The third-order perturbation theory 
thus poses a non-trivial improvement of existing results. It might be
argued \cite{adzhemyan2003} that its severity is probably more than the order of magnitude worse than that
 of the second order. Not only is the number of Feynman diagrams much
 higher, but there are also numerical problems with the correct
 extraction of divergent parts of Feynman diagrams.
 In this paper, we combine both analytical and numerical techniques in order to calculate all necessary renormalization constants.
 We calculate Feynman diagrams numerically and then apply the field-theoretic RG method in an analytical fashion.
 Although some of the partial results  have already been published previously 
%
%
 \cite{mizisin2016,mizisin2018,mizisin2019}, 
%
%
 this paper contains analytical predictions for all critical exponents for the first time.

This paper is organized as follows. In Sec.~\ref{sec:DP} the field-theoretic formulation of DP process is presented.
Renormalization group analysis is performed in Sec~\ref{sec:RG}. Sec.~\ref{sec:com_rg_const} describes non-trivial
aspects of RG calculations in the third order of perturbation theory and Sec.~\ref{sec:DP_class} examines calculated predictions for critical exponents.
 Sec. \ref{sec:conclusion} is reserved for concluding remarks. 
 Supplementary material~\cite{supplement}  contains technical and numerical details about Feynman diagrams, their 
 representation, algebraic structure, and divergent parts. 


\section{Field-theoretic formulation of Directed Percolation}
\label{sec:DP}
{\subsection{Response functional} \label{subsec:action}}

The effective field-theoretic model for directed percolation can be constructed in two different ways, which are completely
 equivalent with respect to universal large-scale behavior.
 The first approach starts with the microscopic version of the DP process on a lattice, which presents a specific reaction-diffusion system.
Initially, the master equation~\cite{Janssen2005} is introduced, whose final form is  reformulated in terms
 of creation and annihilation operators using Doi approach~\cite{Doi1976a,Doi1976b}. Finally, by means of coherent-state path integrals, one ends up with a field-theoretic action~\cite{Tauber2005,Tauber2014,Janssen2005,HHL16}. The second approach is based on phenomenological considerations supplemented by physical insights
  and symmetry considerations~\cite{Janssen2005,Henkel2008}.
   Both of the approaches are very well understood  nowadays~\cite{Janssen2005,Henkel2008,Tauber2014}. Thus we refrain here
from giving a comprehensive derivation and briefly summarize the main aspects of the latter approach.
 
  For DP the fundamental dynamic quantity is the density of active sites $\psi = \psi(t,\mx)$. Universal properties of the DP process in the critical region
  are then effectively captured by the stochastic Langevin-like equation
\begin{equation}
  \partial_t \psi = D_0 \boldnabla^2 \psi + D_0 \tau_0 \psi - \frac{ D_0 g_0 }{2} \psi^2 +  \sqrt{\psi}\zeta,
  \label{eq:stoch_dp}
\end{equation}
where $\zeta=\zeta(t,\mx)$ denotes the noise field to be specified later, $\partial_t = \partial / \partial t$ is
the time derivative, $\boldnabla^2 = \sum_{i=1}^d \partial/\partial x_i^2$ is  the Laplace operator in $d$-dimensions, $D_0$ 
is the diffusion constant, $g_0$ is the coupling constant, and $\tau_0$ measures
 a deviation from the percolation threshold.  For instance, for lattice DP \cite{Henkel2008} the $\tau = p - p_c$ is the deviation from a critical value probability $p_c$ of open bonds, whereas, for laminar-turbulent phase transition \cite{Lemoult2016, Sano2016,Goldenfeld2011}, the $\tau = \mathrm{ Re} - \mathrm{ Re}_c$ is the deviation from a critical value of 
 Reynolds number $\mathrm{ Re }_c$. In general, parameter $\tau$ plays an analogous role to the temperature variable $T - T_c$ in the paradigmatic $\varphi^4 -$theory 
 in critical statics~\cite{Vasilev2004,Zinn2002,Janssen2005}. Factor $1/2$ has been added in front of the $\psi^2$-term for the future convenience.

 RG procedure, which we employ, introduces two different kinds of variables - bare (unrenormalized) quantities and their renormalized counterparts.
 Therefore we denote the former ones with the subscript ``0'', whereas the latter will be written without the subscript ``0''.
 To keep the notation as simple as possible bare functions (e.g. for connected Green functions) will be further 
denoted without the subscript ``0'' and their renormalized versions with the subscript $R$.

From the mathematical point of view, Eq.~\eqref{eq:stoch_dp} should be interpreted in It\^o sense~\cite{Gardiner,Janssen76,Janssen92,Honkonen2012}. 
 In detail, we assume that Gaussian noise field $\zeta (t, \mx)$ has zero mean $\langle \zeta \rangle = 0$ and correlations are given by the relation
\begin{equation}
  \langle \zeta(t, {\mx}) \zeta(t', {\mx}') \rangle = g_0 D_0  \delta (t -t') \delta^{(d)} ({\mx} - {\mx}'),
  \label{eq:stoch_noise}
\end{equation}
where $\delta^{(d)} ({\mx})$ is the $d-$dimensional Dirac delta function. Here, brackets $\langle\cdots\rangle $ denote
  averaging procedure over all possible realizations of the stochastic process.
 The multiplicative character of the noise in Eq.~\eqref{eq:stoch_dp} can be
 regarded as a direct consequence of the absorbing condition, according to which
 all fluctuations have to cease once an absorbing state is entered.

In order to apply the RG method, it is advantageous to recast Langevin formulation~\eqref{eq:stoch_dp}-\eqref{eq:stoch_noise} in terms of functional integrals.
  The stochastic problem \eqref{eq:stoch_dp} can be recast into a quantum-field model with a double set of fields~\cite{Janssen1981,Vasilev2004,Tauber2014}.
 Ensuing De Dominicis-Janssen dynamic functional of the percolation process \cite{deDom76,Janssen76,Janssen2005} takes the following form 
%
%
\begin{align}
  \SA(\varphi) & = \int \dRM x \, \biggl\{ \tpsi(-\partial_t + D_0 \boldnabla^2 + D_0 \tau_0)  \psi \nonumber\\
  & +  \frac{D_0 g_0}{2}  (\tpsi^2 \psi -\tpsi\psi^2) \biggl\},
  \label{eq:action_0}
\end{align}
where for brevity we write $x=(t,\mx)$ and integration measure as $\dRM x = \dRM t \dRM^d x$, $\tpsi$ is an auxiliary Martin-Siggia-Rose response field, and $\varphi\equiv\{ \psi,\tpsi \}$. Let us note
that the response field $\tpsi$ arises from the integration over the noise 
$\zeta$, and it  might be interpreted as the functional Lagrange multiplier~\cite{Vasilev2004,Tauber2014}. 
%
%
 

The existence of an additional symmetry in a quantum field model often provides additional invaluable information that might
 improve the tractability of theoretical analysis and practical calculations. 
 The DP process possesses  symmetry in this sense, which is known in the literature
as rapidity-reversal symmetry~\cite{Hinrichsen2000,Henkel2008}.
It can be directly observed that
the action functional~\eqref{eq:action_0} is invariant
  with respect to the following transformation of fields $\psi$ and $\tpsi$
\begin{equation}
  \psi (t, \mx) \rightarrow - \tpsi (-t, \mx),\quad  \tpsi (t, \mx) \rightarrow -\psi (-t, \mx ).
  \label{eq:symmetry}
\end{equation}
Later on in Sec. \ref{sec:RG}, we employ it to simplify the RG analysis.

{\subsection{Scale invariance}  \label{subsec:scale} } 
 Dynamical
    models are characterized by two independent correlation lengths, besides the spatial 
  length scale $\xi_{\perp}$ there is also temporal length scale $\xi_{\parallel}$ \cite{Vasilev2004,Tauber2014}. 
 In the critical region, we thus expect that the introduced spatio-temporal scales diverge according to some power-law dependence
\begin{equation}
  \xi_{\perp} \sim |\tau_0|^{-\nu_{\perp}}, \quad \xi_{\parallel} \sim |\tau_0|^{-\nu_{\parallel}},
  \label{eq:anis_scaling}
\end{equation}
where $\tau_0$ was introduced in Eq.~\eqref{eq:stoch_dp}.
 Exponents $\nu_{\perp}$ and $\nu_{\parallel}$ take in Eq.~\eqref{eq:anis_scaling} 
 predominantly different values.
 In contrast to relativistic field theories (e.g., quantum electrodynamics), the models
 in statistical physics exhibit strong anisotropic scaling~\cite{Tauber2014}.
 For a given 
 dynamical universality class, it is then convenient to introduce the dynamical exponent $z$ as a ratio
\begin{equation}
   z =  \frac{\nu_{\parallel} }{ \nu_{\perp} }.
   \label{eq:def_exp_z}
\end{equation}  
In the scaling region, the correlation 
 lengths are then simply related as $\xi_{\parallel} \sim \xi_{\perp}^z $.

There are two special values commonly encountered in physics $z=1$ (light-cone spreading)
and $z=2$ (diffusive spreading).
Values $z>2$ correspond to a subdiffusive spreading, whereas $1<z<2$ to a superdiffusive spreading.
 As any other critical exponent $z$ is an universal quantity. 

For the full scaling description of DP process, one additional exponent is needed. Usually, this is the exponent $\beta$, which describes the
 mean particle number in the active phase 
\begin{equation}
  \rho \sim (p_c - p)^\beta \propto (-\tau_0)^\beta.
  \label{eq:def_beta}
\end{equation}

{\subsection{Mean Field approximation} \label{subsec:mean_field} }

Arguably the most naive and direct approach to critical behavior is based on the mean-field (MF) theory, in which spatial variations of order parameter field are neglected~\cite{ChaikinLubensky}. Following this idea and neglecting the noise variable as well, we average Eq.~\eqref{eq:stoch_dp} with an assumption $\langle \psi^2 \rangle \approx \langle \psi \rangle^2$. We 
 immediately obtain an equation for the mean number of active sites $\rho(t) = \langle \psi \rangle$ in the form of an ordinary differential equation
\begin{equation}
  \frac{\dRM \rho}{\dRM t} = D_0
  \left( 
    \tau_0 \rho - \frac{g_0}{2}\rho^2
  \right).
  \label{eq:DP_MF}
\end{equation}

Obviously, there are two stationary solutions to this equation 
$\rho_* = 0$ and $\rho_* =2\tau_0/g_0$. The former corresponds to an
absorbing state and is stable for $\tau_0 <0$. On the other hand, the latter solution
 represents the active state realized for $\tau_0 > 0$.
MF approach thus predicts phase transition at $\tau_0 = 0$. 

From the theoretical point of view, MF theory is justified if the diffusive mixing of particles is much stronger than the influence of
correlations produced by the reactions. However, computer simulations of DP reveal
strong correlation effects in the critical region, and therefore MF is inadequate and cannot yield precise quantitative results~\cite{Henkel2008}.

The spatially amended version of Eq.~\eqref{eq:DP_MF} is simply given by the deterministic partial differential equation
\begin{equation}
  \partial_t \psi = D_0\boldnabla^2 \psi+ D_0
  \left( \tau_0 \psi - \frac{g_0}{2} \psi^2
  \right),
  \label{eq:mean-field}
\end{equation}
where now order parameter field $\psi=\psi(t,\mx)$ depends also on a spatial variable.
Direct scaling analysis~\cite{Henkel2008} yields following values of critical exponents
\begin{equation}
   \quad  \nu_{\perp}^\text{MF} = \frac{1}{2}, \quad
   \nu_{\parallel}^\text{MF} = 1,\quad \beta^\text{MF} = 1.
   \label{eq:MF_exponents}
\end{equation}

{ \section{Renormalization Group Analysis} \label{sec:RG} }
The main aim of the theory of critical behavior~\cite{Zinn2002} is to understand
how the non-trivial macroscopic (large-scale) behavior of the system might emerge from
relatively simple microscopic interactions.
 Related quantitative analysis is usually accompanied by the determination of
  Green functions, i.e., correlation and response functions, as 
functions of the space-time coordinates. In the field-theoretic formulation, these functions can be represented
 in the form of perturbative sums, whose elements are conveniently expressed through
 Feynman diagrams. This formulation provides a convenient theoretical
framework suitable for applying methods of statistical (quantum) field theory, among which  the RG method plays a distinguished role. In particular, it facilitates
 a determination of the infrared (IR) asymptotic (large spatial and time scales) behavior of the Green functions, which is
 of fundamental importance for statistical physics. Feynman diagrams are often plagued with divergences, which have to
 be properly taken care of. In order to eliminate ultraviolet (UV) divergences renormalization procedure has to be applied~\cite{Zinn2002,Vasilev2004}. There are
  various renormalization
prescriptions available, each with its own advantages. In this work, we employ dimensional regularization supplemented with the
 minimal subtraction (MS) scheme. In this scheme, UV divergences manifest themselves
in the form of poles in the small expansion parameters, which are given by a deviation from a critical dimension. A simplifying feature of MS scheme 
is  the neglect of
all finite parts of the Feynman graphs in the calculation of the renormalization constants. In the vicinity of critical points, large fluctuations on all 
spatio-temporal scales dominate the behavior of the system, which in turn results in the IR divergences
in the Feynman graphs. The non-trivial connection between UV and IR divergences~\cite{Vasilev2004,Tauber2014} for logarithmic theory allows us to derive
  RG differential equations, which describe scaling behavior. As a byproduct, an analysis of these equations
  furnishes an efficient calculational technique for critical exponents.

The action Eq.~\eqref{eq:action_0} is  amenable to the field-theoretic methods of quantum field theory \cite{Vasilev2004}.
Green functions correspond to functional 
 averages with respect to the weight functional $\exp(S)$. In practical calculations, it is often more convenient to work preferably with
  connected Green functions.
   These correspond to Feynman diagrams that do not contain disconnected parts~\cite{Zinn2002,Vasilev2004}. They can
  be obtained from the generating functional $\WA$
\begin{equation}
  \WA(A) = \ln \int \DA \varphi \exp \{ \SA (\varphi) + \varphi A \},
  \label{eq:functional}
\end{equation}
by taking the corresponding number of functional derivatives with respect to external sources $A =\{A_{\psi}, A_{\tpsi}\}$ at $A = 0$. Hereinafter, $\varphi$  is the set $\{\psi, \psi'\}$ 
linear term in Eq.~\eqref{eq:functional} is an abbreviated form for the following expression 
\begin{equation}
  \varphi A \equiv \int \dRM x \left[ A_{\psi} (x) \psi (x) + A_{\tpsi}(x) \tpsi(x) \right]. 
  \label{eq:func_scalar}
\end{equation}
  For translationally invariant theories, additional simplification consists in introducing functional $\Gamma (\alpha)$ for 
 one-particle irreducible (1PI) Green functions~\cite{Zinn2002,Vasilev2004}. This is an even more restricted class than that of connected Feynman diagrams, since 1PI
 diagrams are such diagrams that remain connected even after one internal line is cut~\cite{Zinn2002,Vasilev1998}.
 It can be shown that the corresponding functional for 1PI diagrams~\cite{Vasilev2004} can be obtained by means of
 functional Legendre transform with respect to external sources $A$ 
\begin{equation}
  \Gamma (\alpha) = W(A) - \alpha A,  \quad  \alpha (x) = \frac{\delta W(A)}{\delta A(x)}.
  \label{eq:functional_legendre}
\end{equation} 
 The independent argument for the functional $\Gamma$ now becomes a pair of fields
 $\alpha=\{ \alpha_\psi,\alpha_{\tpsi} \}$, while $A = A(\alpha)$ is implicitly determined  by the second equation in Eq.~\eqref{eq:functional_legendre}. 
 The 1PI Green 
functions $\Gamma^{(m,n)}$ are obtained by differentiation with respect to fields $\alpha$
\begin{equation}
  \Gamma^{(m , n)} \equiv 
   \frac{\delta^{m + n} }{\delta \alpha_\psi^m \delta \alpha_{\tpsi}^n } \Gamma (\alpha) \bigg|_{\alpha =0}.
\end{equation}
 As a result of causality, all 1PI Green functions of the form $\Gamma^{(0, n)}$ vanish. The same conclusion applies to related 1PI functions $\Gamma^{(m, 0)}$  as a consequence of rapidity-reversal symmetry \eqref{eq:symmetry}.  

\begin{figure}
\includegraphics[width=7cm]{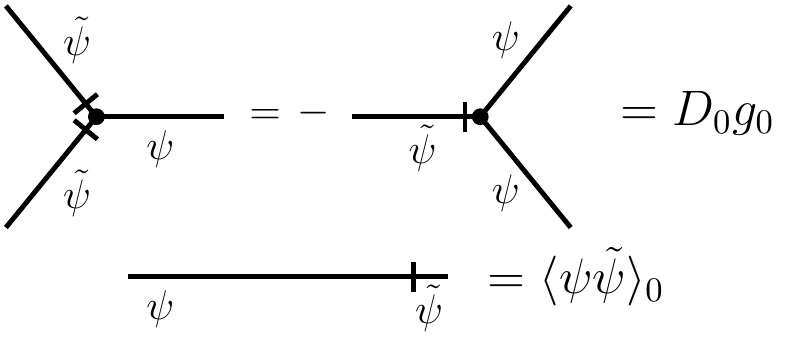}
\caption{Graphical representation of Feynman rules for DP model. The crossed
line corresponds to a response field $\tpsi$. }
\label{fig:DP_rules} 
\end{figure}
To abbreviate the notation, let us relabel variables $\alpha$ according to the prescription $\alpha\rightarrow\varphi$ and obtain
an important relation~\cite{Zinn2002,Vasilev2004} that summarizes the
importance of  
generating functional for 1PI Feynman graphs in a concise manner
\begin{equation}
  \Gamma(\varphi) = \SA(\varphi) +   
  \Gamma_\text{loop}(\varphi),
  \label{eq:RG_equation}
\end{equation}
where $\SA$ is the bare action functional of the model~\eqref{eq:action_0}, and
$ \Gamma_\text{loop}(\varphi)$
 is the sum of all 1PI loop diagrams contributions. 

The starting point of perturbation methods in quantum field theory  is to represent
 Green functions in terms of a corresponding sum of Feynman diagrams, whose
 relevance is controlled by interaction parameters (coupling constants).
 Feynman graphs represent convenient graphical representations 
 of various algebraic structures
that arise from basic perturbation elements, which are formed by propagators and interaction vertices. 

For DP model, we encounter a single bare propagator, which is obtained from the quadratic part of the action functional \eqref{eq:action_0}.
In a frequency-momentum representation, it reads
\begin{equation}
  \langle \psi \tpsi \rangle_0 (\omega, {\mpp}) =\frac{1}{-i\omega + \eps_p},
  \label{eq:prop_omega}
\end{equation}
where $\eps_p = D_0(\mpp^2 + \tau_0)$. In time-momentum representation the propagator
 $\langle \psi \tpsi \rangle_0$ is given by
\begin{equation}  
  \langle \psi \tpsi \rangle_0 (t, {\mpp}) =  \theta(t) \eRM^{- \eps_p t },
  \label{eq:prop_t}
\end{equation}
where $\theta (t)$ is the Heaviside step function.

Non-linear terms in the action~\eqref{eq:action_0} generate two interaction vertices of the model, and they are associated with the vertex factors~\cite{Vasilev2004}
\begin{equation}
  V_{\tpsi\psi\psi} = - V_{\tpsi\tpsi \psi} = D_0 g_0.
  \label{eq:vertex_factors}
\end{equation}   
The Feynman rules for the DP model are depicted in Fig.~\ref{fig:DP_rules} and, in fact, they represent the zero (tree) order of a perturbation theory~\cite{Vasilev2004}. 
Loop diagrams then contribute to higher-order terms with respect to
the coupling constant $g_0$, and
as a rule, they exhibit various divergences in UV or IR sector.
In graphical terms, these statements are part of Eq. \eqref{eq:RG_equation}. 
In particular, the two-point Green function takes the schematic form of the
so-called Dyson equation
\begin{equation}
  \Gamma^{(1,1)} = \langle \psi \tpsi \rangle_{1PI} = -\langle \psi\tpsi \rangle_0^{-1} 
  +\raisebox{-2.8ex}{\includegraphics[width=2.cm]{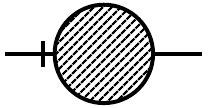}},
  \label{eq:dyson}
\end{equation}
where the shaded diagram corresponds to the loop corrections. Further, for
 the three-point interaction term, we have
\begin{equation}
  \Gamma^{(1,2)} = \langle \psi \psi \tpsi \rangle_{1PI} = V_{\tpsi\psi\psi} + 
  \raisebox{-3.1ex}{\includegraphics[width=2.cm]{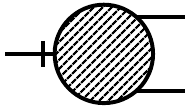}}.
  \label{eq:1PI_vertex}
\end{equation}

The RG technique helps us to eliminate UV divergences, and through a non-trivial connection with IR behavior valid at the 
%
%
upper critical
%
%
 dimension, we can
 extract useful information about
 macroscopic behavior of universal quantities~\cite{Vasilev2004,Tauber2014}.
{\subsection{ Canonical dimensions } \label{subsec:can_dim} }
The starting point of the actual RG analysis is a determination of canonical dimensions for all fields and parameters of the model~\cite{Zinn2002,Vasilev2004}.
As it has been pointed out in Sec.~\ref{subsec:scale},
DP process belongs to dynamical models and for such systems it 
is necessary to introduce two independent 
 dimensions: a frequency dimension $d_{\omega} [Q]$ and
 a momentum dimension $d_k [Q]$, where $Q$ is a given quantity (field or parameter) in the action functional~\eqref{eq:action_0}. 
 We employ standard normalization conditions
\begin{align}
  d_k [k] & = - d_k [x] = 1, & d_k [\omega] & =  d_k [t] = 0, 
  \label{eq:canon1}  
  \\
  d_{\omega} [k] & = d_{\omega} [x] = 0, & d_{\omega} [\omega] & =  - d_{\omega} [t] =1. 
  \label{eq:canon2}
\end{align}
 We aim to study the model  in the asymptotic region $\omega\propto k^2$,  which
 implies that both terms $\partial_t \psi$ and $\boldnabla^2 \psi$ are to be taken
 on an equal footing.
 The total canonical dimension $d_Q$ of a quantity $Q$ is given by a weighted sum of momentum and frequency dimension
\begin{equation}
  d [Q] = d_k [Q] + 2 d_{\omega} [Q].
  \label{eq:canon_dim1}
\end{equation} 
The 
canonical dimensions of all quantities (fields and parameters)
 in the DP action~\eqref{eq:action_0} are listed in Tab.~\ref{tab:canon_dim}.
 
By usual considerations \cite{Vasilev2004}, we derive the formula for
the total canonical 
dimension of an 1PI Green function $\Gamma$ in a form
\begin{equation}
  d [\Gamma^{(m,n)}] = d_k [\Gamma^{(m,n)}] + 2 d_{\omega}[\Gamma^{(m,n)}].
  \label{eq:canon_dim2}
\end{equation} 
We can rewrite this equation 
with the help of Eq.~\eqref{eq:canon_dim1} and Tab.~\ref{tab:canon_dim} 
into a more informative form
\begin{equation}
  d [\Gamma^{(m,n)}] =  d + 2 - n d[\psi] - m d[\tpsi],
  \label{eq:total_dimension}
\end{equation} 
where $n$ and  $m$ are a number of fields $\psi$ and $\tpsi$ entering given Green function $\Gamma^{(m,n)}$. 

As can be seen from Tab.~\ref{tab:canon_dim}, the model
 is logarithmic at space dimension $d_c=4$ when the coupling 
constant $g_0$ is dimensionless.
In the dimensional regularization, it is reasonable to introduce a formally small parameter $\eps$ as a difference
\begin{equation}
  \eps \equiv d_c - d = 4-d .
  \label{eq:def_eps}
\end{equation}
For the logarithmic theory, the total canonical dimension $d[\Gamma]$ represents the formal degree of UV divergence, i.e. 
\begin{equation}  
  \delta_{(m,n)} = d [\Gamma^{(m,n)}]|_{\eps = 0},
  \label{eq:UV_index} 
\end{equation}
  and
  UV divergences appear in a Green function as various poles in $\eps$.
The central step in the RG procedure lies in  removing superficial UV 
 divergences, which may appear only in those 1PI functions $\Gamma$ 
 for which UV index~\eqref{eq:UV_index} $\delta_\Gamma $
  attains non-negative values.
 In a straightforward manner, we determine that for DP model~\eqref{eq:action_0}  the only 1PI functions with UV 
 divergences are  
$\Gamma^{(1,1)}$, $\Gamma^{(2,1)}$ and $\Gamma^{(1,2)}$.

\begin{table}
\caption{Canonical dimensions of the bare fields and bare parameters for the action (\ref{eq:action_0}).}
\begin{ruledtabular}
  \begin{tabular}{ c  c  c  c  c  c }
	$Q$ & $\psi$ & $\tpsi$ & $D_0$ & $\tau_0$ & $g_0$ \\
   \noalign{\smallskip}\hline\noalign{\smallskip}
	$d_{\omega}[Q]$ & $0$ & $0$ & $1$ & $0$ & $0$  \\
	 \noalign{\smallskip}\hline\noalign{\smallskip}
	$d_{k}[Q]$ & $d/2$ & $d/2$ & $-2$ & $2$ & $(4-d)/2$ 
	 \\
	  \noalign{\smallskip}\hline\noalign{\smallskip}
	$d [Q]$ & $d/2$ & $d/2$ & $0$ & $2$ & $(4-d)/2$
  \end{tabular}
\end{ruledtabular}
\label{tab:canon_dim}
\end{table}

Green functions are, in perturbative calculations, naturally ordered
 in terms of a charge $g_0$.
As direct inspection of Feynman diagrams reveals, the 
 actual charge of the perturbation theory is effectively $g_0^2$ rather than $g_0$.
 Therefore it is advantageous to introduce a new charge $u_0$ as
\begin{equation}
  u_0 = g^2_0 
  \label{eq:definition_u}
\end{equation}
with total canonical dimension $d[u_0] =  \eps$.

From the aforementioned considerations and taking into account symmetry \eqref{eq:symmetry}, we infer that
the renormalized action functional for the DP process can be written  in the compact form
%
%
\begin{align}
   \SA_R & = \int\dRM x\, \biggl\{ \tpsi_R(- Z_1 \partial_t + Z_2 D \boldnabla^2 - Z_3 D \tau) \psi_R \nonumber \\
   &+ \frac{D g}{2} Z_4 \mu^{\eps /2} (\tpsi_R^2 \psi_R -\tpsi_R \psi^2_R)\biggl\},
  \label{eq:action_r}
\end{align}
%
%
where $\psi_R,\tpsi_R$ are renormalized fields,
 $\mu$ is the reference mass scale in the MS scheme~\cite{Amit,Zinn2002,Vasilev2004} with dimension $d[\mu]=1$, and $Z_i$ $(i=1,2,3,4)$ 
are renormalization constants that need to be determined in a perturbative fashion.
The fact that the interaction terms enter into the 
renormalization action as a single combination is a consequence of rapidity-reversal symmetry \eqref{eq:symmetry}, i.e., non-linear terms in the renormalized action 
\eqref{eq:action_r} are renormalized with the same renormalization constant $Z_4$. 
Therefore, in the actual calculations, we restrict our attention only to
the 1PI Green function $\Gamma^{(2,1)}$.

Model~\eqref{eq:action_r} is multiplicatively renormalizable~\cite{Janssen1981,Janssen2005,Tauber2014}, 
and all UV divergences can be absorbed by the following
renormalization prescription
\begin{align}
  \psi & = Z_{\psi} \psi_R, & \tpsi & = Z_{\tpsi} \tpsi_R,
  \label{eq:orig_RG1}
  \\
   \tau_0 & = Z_{\tau} \tau + \tau_c, & g_0 & = \mu^{\eps/2} Z_g g, & D_0 & = Z_D D,
   \label{eq:orig_RG2}
\end{align}
where $Z_i$ $(i\in\{\psi,\tpsi,\tau,g,D \})$ are the corresponding renormalization constants.
 The term $\tau_c$ denotes the additive renormalization contribution (like a shift of
 a critical temperature $T_c$ in the theory of critical behavior).
Renormalization of the effective charge $u$ directly follows from its definition~\eqref{eq:definition_u}
\begin{equation}
  u_0  = \mu^{\eps} Z_u u, \quad Z_u=Z_g^2.
  \label{eq:renorm_u}
\end{equation}
Clearly, renormalized charge $u$ is dimensionless.

 Relations between the renormalization constants for fields and parameters and
 renormalization constants $Z_i;i=1,2,3,4$ can be deduced straightforwardly from the renormalized action \eqref{eq:action_r}
\begin{align}
   Z_1 & = Z_{\psi} Z_{\tpsi},\\
    Z_2 & =  Z_D Z_{\psi} Z_{\tpsi},\\
    Z_3 & = Z_{\tau} Z_D Z_{\psi} Z_{\tpsi}, \\
   Z_4 & = Z_{g} Z_D Z_{\psi}^2 Z_{\tpsi}  =  Z_{g} Z_D Z_{\psi} Z_{\tpsi}^2.
\end{align}
Inverting these relations yields
\begin{align}
   Z_{\psi} & = Z_{\tpsi} = Z_1^{1/2},
  \label{eq:rg_relation2a}   
   \\
   Z_D & = Z_2 Z_1^{-1}, 
   \label{eq:rg_relation2b}      
    \\   
   Z_{u} & 
   = Z_g^2= Z_4^2 Z_2^{-2} Z_1^{-1},
  \label{eq:rg_relation2c}      
    \\
   Z_{\tau} & = Z_3 Z_2^{-1},
  \label{eq:rg_relation2d}
\end{align}
where the first equality is a consequence of symmetry \eqref{eq:symmetry}.

In actual calculations, we thus need to concentrate solely on an analysis of renormalization constants $Z_i$, where $i=1,2,3,4$.
{\section{Computation of RG Constants} \label{sec:com_rg_const}}
In this section, we review major points in a numerical determination of RG constants. A summary of particular algebraic details on Feynman 
diagrams, such as symmetry factors and their pole parts can be found in the supplementary material~\cite{supplement}. 

In the employed MS scheme, renormalization constants takes the following general form
\begin{equation}
  Z_i (u, \eps) = 1 + \sum\limits_{k=1}^{\infty} u^k \sum\limits_{l=1}^k \frac{c_{i,kl}}{\eps^{l}},  
  \label{eq:general_Z_MS}
\end{equation}
where $c_{i,kl}$ are pure numerical factors (they do not depend on any other model parameters such as temperature variable $\tau$, charge $u$, or diffusion
 constant $D$~\cite{Vasilev2004}). 

%
%
 Let us note that it is convenient to
  absorb a common geometric factor into a redefinition of the coupling constant $u$ in the following way
%
%
\begin{equation}
  u \frac{S_d}{2(2\pi)^d} \rightarrow u, \quad S_d = \frac{2\pi^{d/2}}{\Gamma(d/2)},
  \label{eq:redefinition}
\end{equation}
where $S_d$ is the surface of the unit sphere in the $d$-dimensional space and $\Gamma(x)$ 
denotes Euler's gamma function.
%
%
In what follows such rescaling is always implied.
%
%

In general, counterterms of 1PI functions $\Gamma^{(m,n)}$ have the form of polynomials in external momenta, frequencies, and mass parameter 
$\tau_0$, respectively.
 The corresponding degree of divergence for a Green function $\Gamma^{(m,n)}$ is determined by the index 
 $\delta_{(m,n)}$ 
 obtained easily from Eqs.~\eqref{eq:total_dimension} 
 and \eqref{eq:UV_index}.
  For the function $\Gamma^{(1,1)}$ we get 
  $\delta_{(1,1)} = 2$, from which we conclude that its counterterms is necessarily
  proportional to external frequency  $\omega,$ inflowing momentum $ \mpp^2$, 
  and temperature parameter $\tau_0$. On the other hand, for function $\Gamma^{(2,1)}$
  we have 
  $\delta_{(2,1)} = 0$, and its  counterterms could not depend on any external momentum or frequency, or mass $\tau_0$, respectively.
  
 To eliminate divergences in the theory, it suffices to eliminate them from the following quantities


\begin{align}
\Gamma_1 & = \partial_{i\omega} \Gamma^{(1,1)} \bigg|_{\mpp = 0, \omega=0}, 
\label{eq:Gamma1} \\
\Gamma_2 & = - \frac{1}{2 D_0} \partial_{p}^2 \Gamma^{(1,1)} \bigg|_{\mpp = 0, \omega=0}, 
\label{eq:Gamma2}\\
\Gamma_3 & = - \frac{1}{D_0} \partial_{\tau_0} \Gamma^{(1,1)} \bigg|_{\mpp=0, \omega=0}, 
\label{eq:Gamma3} \\
\Gamma_4 & =  - \frac{1}{g_0 D_0}  \Gamma^{(2,1)} \bigg|_{\mpp=0, \omega=0} .
\label{eq:Gamma4} 
\end{align}

 

These quantities $\Gamma_i;i=1,2,3,4 $ are dimensionless and normalized in such a way that $\Gamma_i |_{u_0 = 0} = 1$. 
%
%
 They can depend only on the dimensionless ratio $u_0 / \tau_0^{\varepsilon /2}$. In a straightforward manner, it is possible to pull out
 a factor $\tau_0^{-\varepsilon n /2} $ in expressions \eqref{eq:Gamma1}-\eqref{eq:Gamma4} by performing a rescaling
 of an internal momentum $\mk$ according to
 the prescription $\mk \rightarrow \mk / \tau_0^{1/2}$. 
 The resulting perturbation expansion then takes the form
%
%

\begin{equation}
\Gamma_i (\tau_0, u_0) = 1 + \sum\limits_{n=1} (-u_0)^n \tau_0^{-\varepsilon n /2} \Gamma_i^{(n)},
\label{eq:LA1}
\end{equation}
where $\Gamma_i^{(n)}$ are fully dimensionless quantities. 
 Effectively this corresponds to a calculation of the expansion
  coefficients $\Gamma^{(n)}_i$ at $\tau_0 = 1$.

RG theory \cite{Vasilev2004,Zinn2002} leads to an important relation between renormalized and bare 1PI Green functions
\begin{equation}
  \Gamma^{(m,n)}_R (\ldots; e, \mu)  =
  Z_{\tpsi}^m Z_{\psi}^n 
  \Gamma^{(m,n)} 
  \left( \ldots; e_0(e,\mu) \right),
  \label{eq:RG_relation2}
\end{equation}
where $\ldots$ indicates common frequency-momentum dependence $\{\omega, \mk \}$, $e=\{ D,\tau,u\}$ is a set of renormalized parameters and $e_0=\{ D_0,\tau_0,u_0\}$ a corresponding set of bare counterparts. Using Eq.~\eqref{eq:RG_relation2} we immediately get for the renormalized analogs of the quantities 
  \eqref{eq:Gamma1}-\eqref{eq:Gamma4}
 useful functional relations
\begin{equation}
\Gamma_{iR} = Z_i \Gamma_i \left( \tau Z_{\tau}, \mu^{\varepsilon} u Z_u \right);
i=1,2,3,4.
\label{eq:LA2}
\end{equation}

Taking into account Eq. \eqref{eq:LA1}, we further derive
\begin{align}
\Gamma_{iR} (\tau, \mu, u) & = Z_i \biggl[ 1 + \sum\limits_{n=1} Z_u^n (-u)^n  \left( \frac{\mu^2}{\tau } \right)^{n \varepsilon/2} \nonumber\\
& \times Z_{\tau}^{- n \varepsilon/2} \Gamma_i^{(n)} \biggl] .
\label{eq:LA3}
\end{align}
The renormalization constants have to be chosen so that the right-hand side 
of Eq. \eqref{eq:LA3}  does not contain poles in $\varepsilon$. It is
 well known from the renormalization theory \cite{Zinn2002,Vasilev2004} that renormalization constants $Z_i$ in the scheme MS do not
  depend on the renormalization mass $\mu$. Hence, an identical requirement can be imposed on the quantity
\begin{equation}
\Gamma_{iR} (\tau = \mu^2, u) = Z_i \left[ 1 + \sum\limits_{n=1} Z^n (-u)^n \Gamma_i^{(n)} \right], 
\end{equation}
where for brevity, we have introduced the abbreviation
\begin{equation}
  Z \equiv 
   Z_u Z_{\tau}^{- \varepsilon/2} .
\end{equation}

The final expressions for RG constants $Z_i $ can be additionally verified by direct substitution of them into  \eqref{eq:LA3}  and demonstrating
 that the pole terms containing non-analytic expressions like  $\ln (\mu^2 /\tau)$
 cancel out \cite{Vasilev2004}.

\subsection{Renormalization constants}
\label{subsec:rg_constants}

\begin{table}
\caption{The number of Feynman diagrams for DP model that need to be analyzed 
 for the three-loop approximation. }
\begin{ruledtabular}
\begin{tabular}{c  c  c  c}
Diagrams & $1$-loop & $2$-loop & $3$-loop \\
 \noalign{\smallskip}\hline\noalign{\smallskip}
  $\langle \tpsi \psi \rangle$ & $1$ & $2$ & $17$ \\
   \noalign{\smallskip}\hline\noalign{\smallskip}
  $\langle \tpsi \psi \psi \rangle$ & $1$ & $11$ & $150$ \\
\end{tabular}
\label{tab:dp_loop}
\end{ruledtabular}
\end{table}

The Green function $\Gamma^{(1,1)}$  consists of the following Feynman diagrams,
 which we group according to the number of loops.
The one-loop contribution includes a single Feynman diagram
\begin{align}
  \frac{1}{2} \raisebox{-2.25ex}{\includegraphics[width=2.cm]{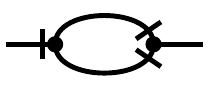}}.
  \label{eq:self_energy1}
\end{align}
Hereinafter, the numerical factor in front of a diagram corresponds to its symmetry factor~\cite{Vasilev2004}.
Further, there are two two-loop diagrams
\begin{align}  
    \frac{1}{2} \raisebox{-1.8ex}{\includegraphics[width=\widthprop]{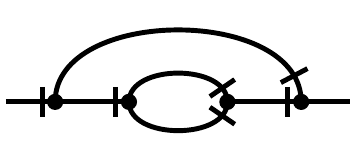}}
   +  \raisebox{-3.2ex}{\includegraphics[width=\widthprop]{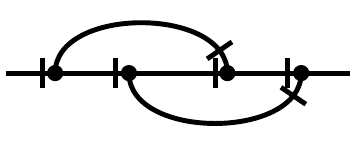}},
\label{eq:self_energy2}
\end{align}
and, finally, the most important contributions to our aim are summarized by a
following sum of three-loop Feynman diagrams
\begin{align}  
 & \quad \raisebox{-0.2ex}{\includegraphics[width=\widthprop]{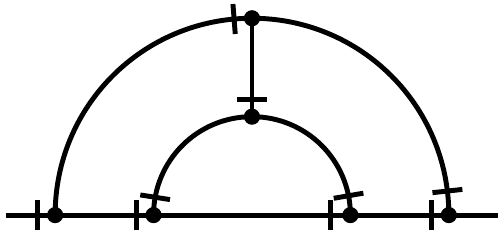}} 
   +  \raisebox{-0.2ex}{\includegraphics[width=\widthprop]{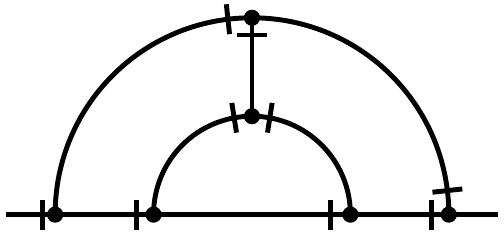}} \nonumber  \\
 &  +\raisebox{-0.2ex}{\includegraphics[width=\widthprop]{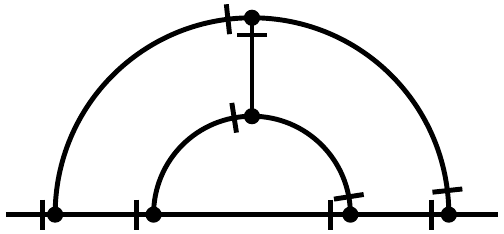}} 
  + \raisebox{-0.2ex}{\includegraphics[width=\widthprop]{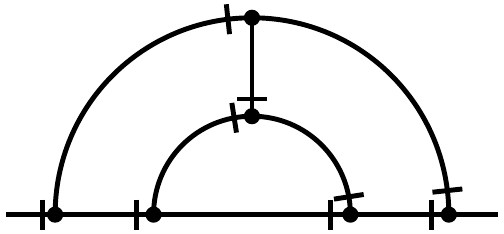}}  \nonumber  \\
  & +  \raisebox{-3.75ex}{\includegraphics[width=\widthprop]{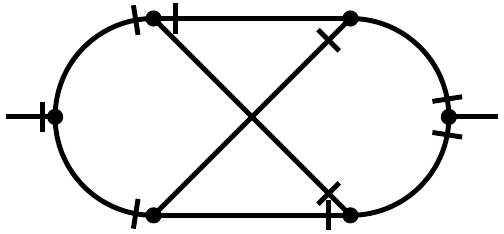}}
   + \frac{1}{4} \raisebox{-3.75ex}{\includegraphics[width=\widthprop]{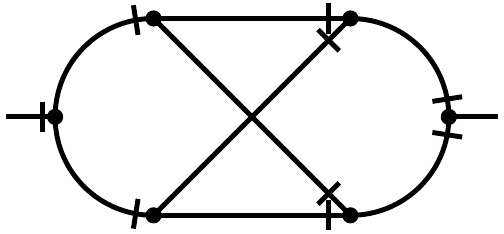}}  \nonumber  \\
  & +  \frac{1}{8} \raisebox{-6.ex}{\includegraphics[width=\widthprop]{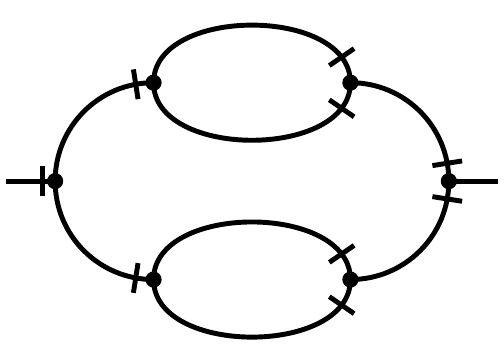}}
   +  \frac{1}{2} \raisebox{-1.ex}{\includegraphics[width=\widthprop]{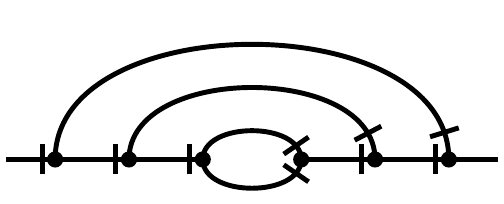}} \nonumber  \\
 & + \frac{1}{2} \raisebox{-4.25ex}{\includegraphics[width=\widthprop]{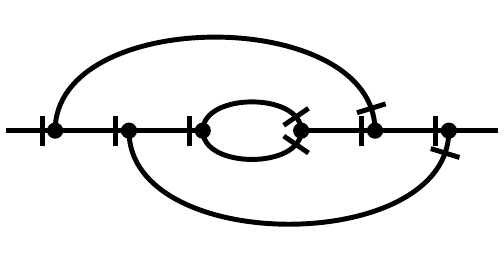}}
   + \raisebox{-3.75ex}{\includegraphics[width=\widthprop]{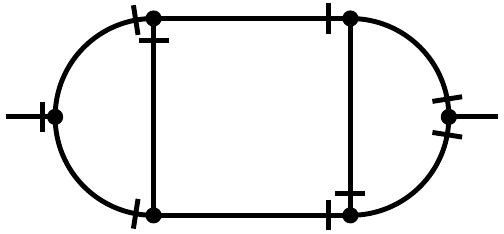}}  \nonumber \\
  & +  \raisebox{-3.5ex}{\includegraphics[width=\widthprop]{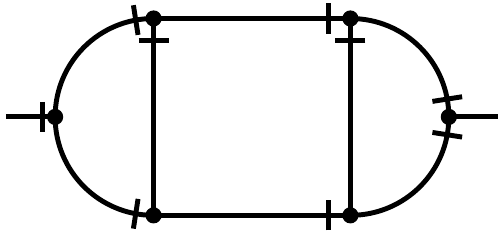}}
   + \frac{1}{2} \raisebox{-1.ex}{\includegraphics[width=\widthprop]{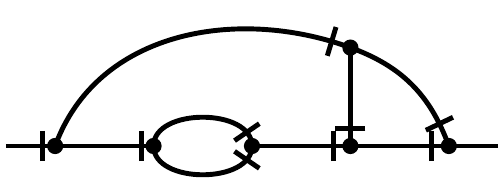}} \nonumber  \\
 & + \frac{1}{2} \raisebox{-1.ex}{\includegraphics[width=\widthprop]{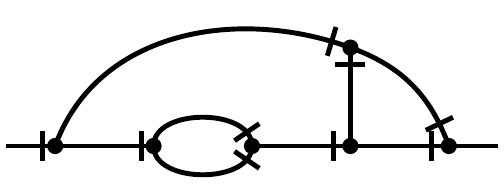}} 
  +  \frac{1}{2} \raisebox{-1.ex}{\includegraphics[width=\widthprop]{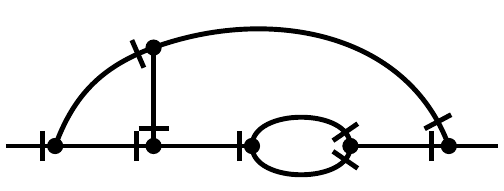}} \nonumber  \\
 &  + \frac{1}{2} \raisebox{-1.ex}{\includegraphics[width=\widthprop]{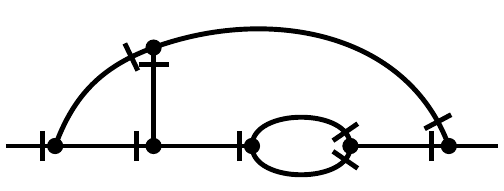}} 
  + \frac{1}{2} \raisebox{-1.ex}{\includegraphics[width=\widthprop]{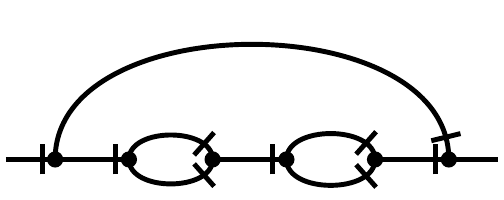}} \nonumber  \\
  & +  \raisebox{-2.ex}{\includegraphics[width=\widthprop]{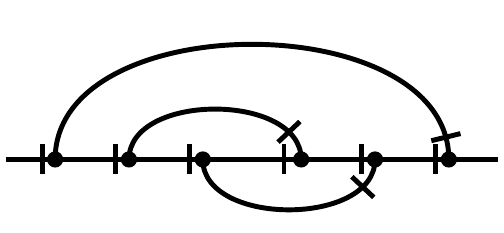}}.
\label{eq:self_energy3}
\end{align}

The function $\Gamma^{(1,2)}$ in the one-loop approximation corresponds to  a single triangle-like Feynman graph
\begin{equation}
 \raisebox{-1.3ex}{\includegraphics[width=2.3cm]{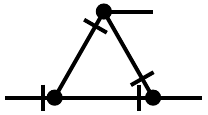}}.
  \label{eq:vertex_green1} 
\end{equation} 
In the two-loop approximation, there are altogether three distinct graph topologies
(skeleton diagrams)
\begin{align}  
  &\raisebox{-5.5ex}{\includegraphics[width=\widthvertex]{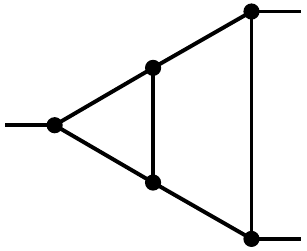}}^{(\times 6)} 
   + &  \raisebox{-5.5ex}{\includegraphics[width=\widthvertex]{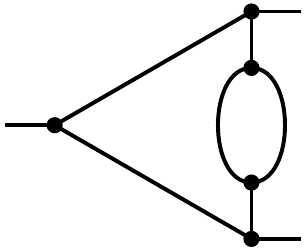}}^{(\times 3)} \nonumber  \\
  + & \raisebox{-5.5ex}{\includegraphics[width=\widthvertex]{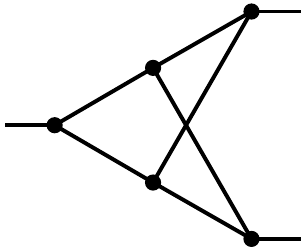}}^{(\times 2)},
  \label{eq:vertex_green2} 
\end{align}  
where to make the notation more compact we have not explicitly denoted response fields
 as introduced in graphical rules from Fig. \ref{fig:DP_rules}.
 Instead, we have explicitly written a numerical factor after a diagram, which corresponds to its multiplicity.
 For a given diagram, it is then straightforward to reconstruct needed configurations
 of response field $\tpsi$.
For example, the second graph at the two-loop approximation \eqref{eq:vertex_green2} corresponds to
\begin{align}
\raisebox{-5ex}{\includegraphics[width=\widthvertex]{fig_three_2_loop_B.pdf}}^{(\times 3)}  =  
\raisebox{-4.ex}{\includegraphics[width=\widthvertex]{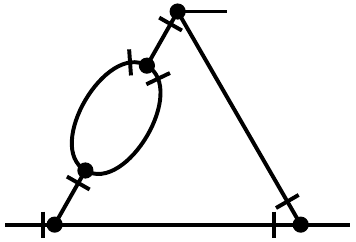}}
\nonumber \\
+ \raisebox{-4.3ex}{\includegraphics[width=\widthvertex]{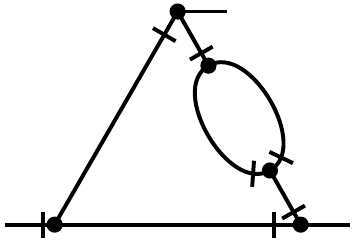}}
+ \raisebox{-5.5ex}{\includegraphics[width=\widthvertex]{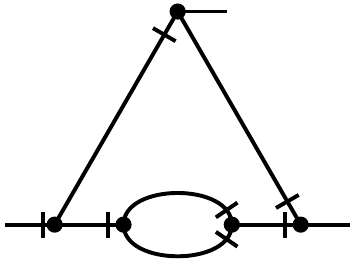}} .
\label{eq:diag_2_loop} 
\end{align} 

  Last, the three-loop approximation comprises the following graph topologies
\begin{align}   
  & \quad 
  \raisebox{-6.ex}{\includegraphics[width=\widthvertex]{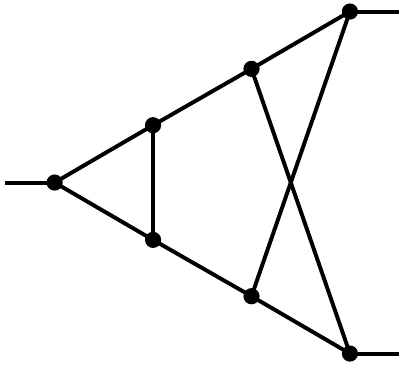}}^{(\times 8)}+
  \raisebox{-6.ex}{\includegraphics[width=\widthvertex]{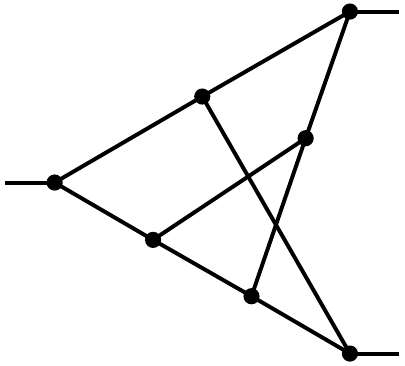}}^{(\times 5)} \nonumber  \\
 &  + \raisebox{-6.ex}{\includegraphics[width=\widthvertex]{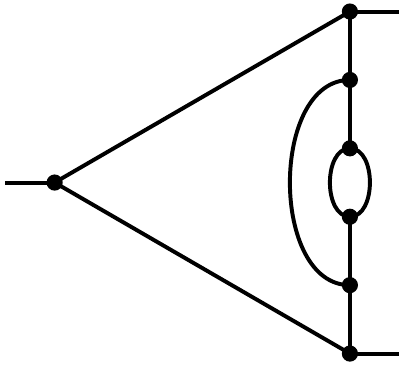}}^{(\times 3)}
 + \raisebox{-6.ex}{\includegraphics[width=\widthvertex]{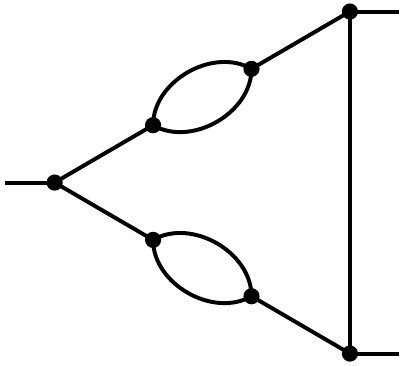}}^{(\times 3)}   \nonumber  \\
 &+  \raisebox{-6.ex}{\includegraphics[width=\widthvertex]{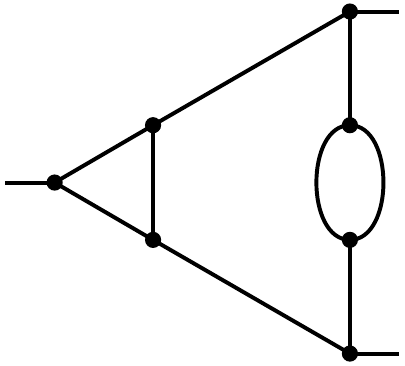}}^{(\times 6)}
  + \raisebox{-6.ex}{\includegraphics[width=\widthvertex]{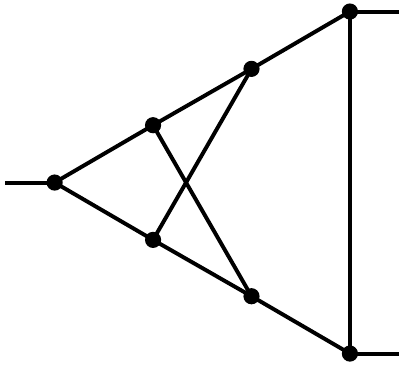}}^{(\times 9)}  \nonumber  \\
  & + \raisebox{-6.ex}{\includegraphics[width=\widthvertex]{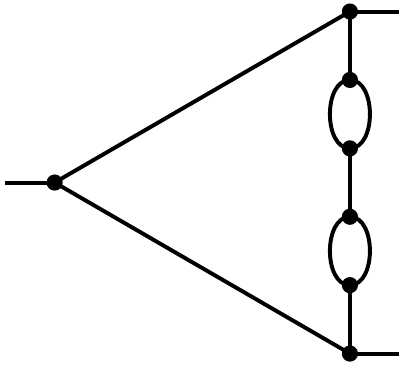}}^{(\times 3)}
   +  \raisebox{-6.ex}{\includegraphics[width=\widthvertex]{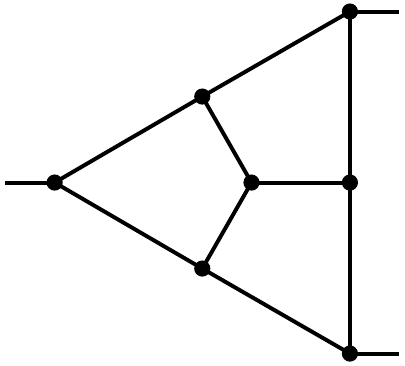}}^{(\times 6)} \nonumber  \\
 & + \raisebox{-6.ex}{\includegraphics[width=\widthvertex]{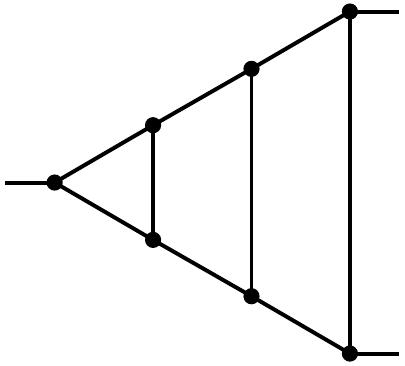}}^{(\times 12)} 
  + \raisebox{-6.ex}{\includegraphics[width=\widthvertex]{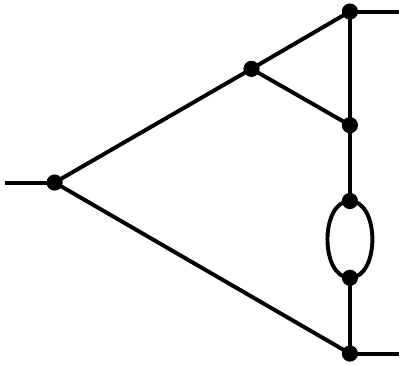}}^{(\times 12)} 
   \nonumber  \\
 &+  \raisebox{-6.ex}{\includegraphics[width=\widthvertex]{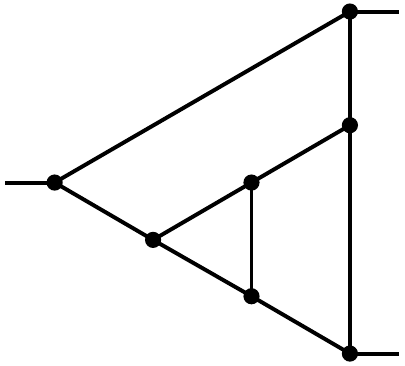}}^{(\times 24)} 
  + \raisebox{-6.ex}{\includegraphics[width=\widthvertex]{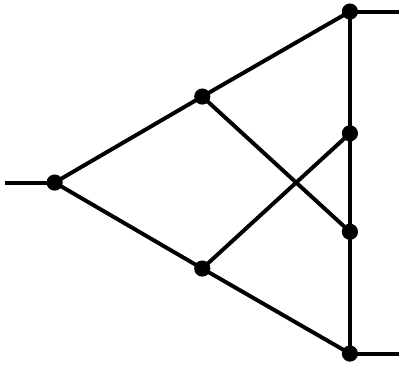}}^{(\times 18)}  \nonumber  \\
  &+ \raisebox{-6.ex}{\includegraphics[width=\widthvertex]{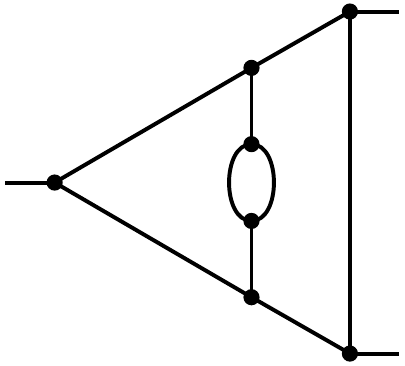}}^{(\times 6)}
   +  \raisebox{-6.ex}{\includegraphics[width=\widthvertex]{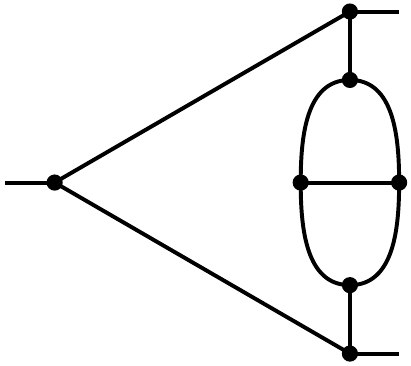}}^{(\times 3)} \nonumber  \\
 & + \raisebox{-6.ex}{\includegraphics[width=\widthvertex]{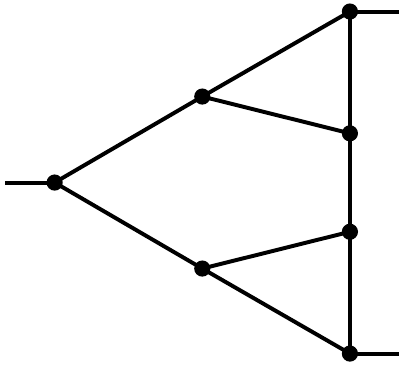}}^{(\times 12)}
  +  \raisebox{-6.ex}{\includegraphics[width=\widthvertex]{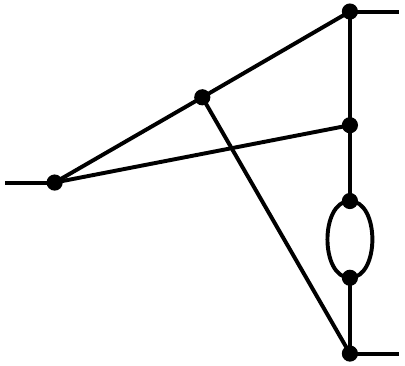}}^{(\times 8)}  \nonumber  \\
 & +  \raisebox{-6.ex}{\includegraphics[width=\widthvertex]{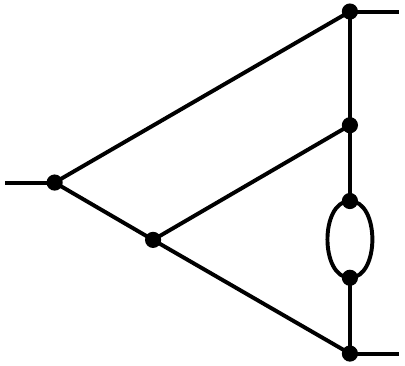}}^{(\times 12)}.
  \label{eq:vertex_green3} 
\end{align}

A computer algorithm for the calculation of Feynman diagrams is similar to that of model A~\cite{Adzhemyan2018}, it generalizes the original Sector Decomposition algorithm~\cite{Binoth2000} to the models of critical dynamics.
For actual numerical calculations, we have  applied the Vegas algorithm from 
  CUBA library with a new implementation \cite{Hahn2005}. Results for coefficients  $\Gamma_i^{(k)}$ read
 
\begin{align}
    \Gamma_1^{(1)}& = \frac{1}{8\eps}-\frac{1}{16}+\frac{\pi^2\eps}{192},
    \\
    \Gamma_1^{(2)}& = \frac{7}{128\eps^2}-0.053082575(5)\frac{1}{\eps}+0.041287893(4),\quad\\
    \Gamma_1^{(3)} & = \frac{91}{3072\eps^3}-0.0359754(2)\frac{1}{\eps^2}+0.0226904(4)\frac{1}{\eps},\\
    \Gamma_2^{(1)}& =\frac{1}{16\eps}-\frac{1}{32}+\frac{\pi^2\eps}{384},\quad\\
    \Gamma_2^{(2)}& = \frac{13}{512\eps^2}-0.02594616442(3)\frac{1}{\eps}+0.020722022(2),
    \\
    \Gamma_2^{(3)}&=\dfrac{325}{24576\eps^3}-0.01710044(3)\frac{1}{\eps^2}+0.01059489(8)\frac{1}{\eps},\quad\\
    \Gamma_3^{(1)}& = \frac{1}{4\eps}-\frac{1}{8}+\frac{\pi^2\eps}{96},\\
    \Gamma_3^{(2)}&=\frac{1}{8\eps^2}-\frac{7}{64\eps}+0.07874750(3),\quad\\
    \Gamma_3^{(3)}& = \frac{325}{24576\eps^3}-0.0765338(3)\frac{1}{\eps^2}+0.0430058(6)\frac{1}{\eps},\\
    \Gamma_4^{(1)}&=\frac{1}{2\eps}-\frac{1}{4}+\frac{\pi^2\eps}{48},\quad\\
    \Gamma_4^{(2)}& = \frac{5}{16\eps^2}-\frac{1}{4\eps}+0.17227563(2),\\
    \Gamma_4^{(3)}& = \frac{5}{24\eps^3}-0.1921389(7)\frac{1}{\eps^2}+0.098458(2)\frac{1}{\eps}.
\end{align}
From these expressions, we obtain RG constants $Z_1 - Z_4$ to the
three-loop approximation in a straightforward manner
\begin{widetext}
\begin{align}
		Z_1 & = 1 +\frac{u}{8\eps} + u^2 \bigg[\frac{7}{128\eps^{2}} 
		- 0.013323675(6)\frac{1}{\eps} \bigg]  + u^3 \bigg[\frac{91}{3072\eps^{3}} - 0.024562(2)\frac{1}{\eps^{2}} 
		+0.0051821(5) \frac{1}{\eps} \bigg],
	\label{eq:Z1}		
		\\ 
Z_2 & =  1 + \frac{u}{16\eps}  + u^2 \bigg[\frac{13}{512\eps^{2}} 
   -  0.0053038358(4)\frac{1}{\eps}  \bigg]  + u^3\bigg[\frac{325}{24576\eps^{3}} - 0.01110090(3)\frac{1}{\eps^{2}} 
   + 0.00072908(8) \frac{1}{\eps} \bigg],
	\label{eq:Z2}   
   \\ 
Z_3 & =  1 +\frac{u}{4\eps}+ u^2  \bigg[\frac{1}{8\eps^{2}} 
   -  \frac{5}{128\eps} \bigg] 
    + u^3\bigg[\frac{7}{96\eps^{3}} - 0.0617665(3)\frac{1}{\eps^{2}} 
   + 0.0253140(6) \frac{1}{\eps} \bigg],
	\label{eq:Z3}   
   \\ 
Z_4 & =  1 +  \frac{u}{2\eps} + u^2  \bigg[\frac{5}{16\eps^{2}} 
   -  \frac{7}{64\eps} \bigg]  + u^3 \bigg[\frac{5}{24\eps^{3}} - 0.1743138(7)\frac{1}{\eps^{2}} 
   + 0.0888244(15) \frac{1}{\eps} \bigg]. 
   	\label{eq:Z4}
\end{align}
\end{widetext}
 Complete diagram-by-diagram results are summarized in supplementary materials~\cite{supplement}. 
%
%
 Feynman diagrams were encoded using efficient scheme
  based on so-called Nickel index~\citep{Nagle1966,Nickel1977,Batkovich2014}.
 In Appendix, we also give
 additional relations between diagrams that were used as an additional crosscheck.

%
%
In a straightforward manner, it is possible to calculate
corresponding expressions for renormalization constants 
for parameters and fields introduced in Eqs. \eqref{eq:orig_RG1}-\eqref{eq:orig_RG2}.
They explicit forms read
\begin{widetext}
\begin{align}
Z_{\psi} & = 1 + \frac{u}{16 \eps} + u^2 \left[0.02539062504(4)\frac{1}{\eps^2} - 0.006661838(3) \frac{1}{\eps} \right]  + u^3 \biggl[0.01322428(3) \frac{1}{\eps^3} - 0.01186460(8) \frac{1}{\eps^2}  \nonumber\\
& + 0.0025911(2) \frac{1}{\eps}\biggl], \\
Z_{D} & = 1 - \frac{u}{16 \eps} + u^2 \left[- 0.02148437507(9) \frac{1}{\eps^2} + 0.008019839(5) \frac{1}{\eps}\right] + u^3 \biggl[- 0.01029458(5) \frac{1}{\eps^3} + 0.0116260(2) \frac{1}{\eps^2} \nonumber\\
& - 0.0044531(5) \frac{1}{\eps}\biggl], \\
Z_{\tau} & = 1 + \frac{3 u}{16 \eps} + u^2 \left[0.0878906250(2) \frac{1}{\eps^2} - 0.033758658(11) \frac{1}{\eps}\right] + u^3 \biggl[0.04943859(6) \frac{1}{\eps^3} - 0.0475612(2) \frac{1}{\eps^2}
\nonumber \\ 
& + 0.0245840(6) \frac{1}{\eps}\biggl], \\
Z_u & = 1 + \frac{3 u}{4 \eps} + u^2 \left[ \frac{9}{16\eps^2} - 0.19481863(3)\frac{1}{\eps} \right]
+ u^3 \left[ 0.4218749(3) \frac{1}{\eps^3} - 0.3409311(13) \frac{1}{\eps^2} + 0.171009(3) \frac{1}{\eps}\right].
\end{align}
\end{widetext}
%
%

{\subsection{RG equation} \label{subsec:RG_eq}}
We have been able to determine RG constants in the MS scheme up to the third-order in perturbation theory. In the calculation process itself,
we have used various verification procedures in order to avoid any conceivable
errors. 
All relevant information regarding each of the
three-loop Feynman diagrams can be found in supplementary material~\cite{supplement}.

Once the calculation of RG constants is successfully accomplished, we are in a position to analyze the asymptotic behavior of experimentally measurable quantities, in 
particular, critical exponents. These govern scaling properties of the DP model~\cite{Hinrichsen2000,Henkel2008} in
 the critical region $\tau \rightarrow 0$. 
Although many permissible Green functions might be studied, here, we focus on the simplest non-trivial functions. As a result, we obtain
numerical predictions for critical exponents.
Additional work would be needed for a calculation of universal moment ratios~\cite{Henkel2008}, and
 a much more demanding task would be a RG analysis of so-called scaling functions~\cite{Lubeck2004}.

In order to derive  basic RG differential equations, let us introduce a differential
operator
\begin{equation}
  \tilde{D}_{\mu} \equiv \mu \partial_{\mu}\bigg|_0, 
\end{equation}
where $\partial_{\mu}|_0$ is the corresponding derivative at fixed bare parameters.
Applying it to Eq.~\eqref{eq:RG_relation2} yields the basic RG differential equations~\cite{Amit,Vasilev2004} for renormalized 1PI Green functions 
\begin{equation}  
  \left(\tilde{D}_{\mu} - n \gamma_{\psi} -  m \gamma_{\tpsi}\right)
  \Gamma^{(m,n)}_R ( \ldots; e, \mu) = 0. 
\end{equation}
For the DP model, the differential operator $ \tilde{D}_\mu $ 
can be expressed through renormalized variables as follows
\begin{equation}
  \tilde{D}_{\mu} = \mu \partial_{\mu} + \beta_{u}\partial_{u} - \tau \gamma_{\tau} \partial_{\tau} - D \gamma_D \partial_D.
\end{equation}
Here, $\beta_u$ is the beta function of the
 coupling constant $u$, and $\gamma_Q;Q\in\{\psi,\tpsi,D,\tau\}$
 are anomalous dimensions for fields and parameters. 
 They are defined~\cite{Zinn2002,Vasilev2004} as 
\begin{equation}
  \gamma_Q \equiv \mu \partial_{\mu}\bigg|_{0} \ln Z_Q.
  \label{eq:ad_def1}
\end{equation}
 In a straightforward fashion, this relation can be rewritten into a more useful
 form
\begin{equation}
  \gamma_Q  = - \eps \frac{u \partial_u \ln Z_Q}{1 + u\partial_u \ln Z_u},
  \label{eq:ad_def2}
\end{equation}
 by which the anomalous dimension is expressed solely in terms of renormalized variables.
  The beta function $\beta_u$ governs the RG flow of the theory, and DP process is an example of the single charge theory.
  Hence, we only have to deal with one beta function   
\begin{equation}
  \beta_{u} \equiv \mu \partial_{\mu}\bigg|_{0} u. 
  \label{eq:beta_def1}
\end{equation}
Using similar considerations as for an anomalous dimension in Eq.~\eqref{eq:ad_def2} and
relation \eqref{eq:renorm_u} we can express the beta function $\beta_u$
in two different ways
\begin{align}
\beta_{u} & =  -u (\eps + \gamma_u) \label{eq:beta_def2} \\
          &= -\eps \frac{u}{ 1 + u \partial_u \ln Z_u}. 
\label{eq:beta_def3}
\end{align}

Final predictions for the anomalous  dimensions take the following form
\begin{align}
  \gamma_{\psi}  = & -\frac{u}{16} + 0.013323675(5) u^2 - 0.0077732(7) u^3,
  \label{eq:gamma_psi} \\
  \gamma_{D}  = & \frac{u}{16} - 0.016039678(11) u^2 + 0.0133592(13) u^3,
     \label{eq:gamma_D} \\
  \gamma_{\tau} = & -\frac{3 u}{16} + 0.06751732(2) u^2 - 0.073755(2) u^3,
    \label{eq:gamma_tau} \\
  \gamma_u  = & -\frac{3 u}{4} + 0.38963725(5) u^2 - 0.513026(9) u^3,
    \label{eq:gamma_u}
\end{align}
where the number in brackets corresponds to a numerical error stemming from applied integration procedures. 
As can be directly seen in expressions~\eqref{eq:gamma_psi}-\eqref{eq:gamma_u}, the
obtained results display a high level of accuracy.

Finally, the beta function $\beta_u$ takes in the three-loop approximation form
\begin{align}
 \beta_u 
  & = - u \bigg( \eps -\frac{3 u}{4} +  0.38963725(5) u^2 
   \nonumber\\  
   & - 0.513026(9) u^3 \bigg).
  \label{eq:beta_u}
\end{align}
Expressions~\eqref{eq:gamma_psi}-\eqref{eq:beta_u} can be regarded as the most important results of this paper. 
To the best of our knowledge, this is the first time
field-theoretical 
predictions for three-loop corrections are presented for a non-equilibrium model.

{\section{Directed Percolation Universality Class} \label{sec:DP_class}}
The existence of an IR attractive fixed point ensures the scaling behavior of theory~\cite{Amit,Vasilev2004}. 
The asymptotic behavior is then governed by IR attractive fixed points of corresponding 
RG equations. 
The coordinates of fixed points can be found from the requirement that all beta
functions of a model simultaneously vanish 
\begin{equation}
  \beta_i(g^*) = 0,\quad \beta_i = \mu \partial_\mu \biggl|_0 g_i,
\end{equation}
where $g_i$ are charges of the theory. IR stability of fixed points is determined from a matrix $\Omega$ of the first
 derivatives with respect to coupling constants $\Omega_{ij} = {\partial \beta_{g_i}/ \partial g_j}$. 
 A fixed point is IR attractive when all eigenvalues of $\Omega$ are positive. 
 
 For DP process, a stability analysis is straightforward as we have to deal with a single  beta function $\beta_u$.
 Hence, to determine the stability of given fixed points only
 one derivative $\partial \beta_u / \partial u$ is needed.
   From the standard requirements imposed on an IR attractive fixed point,
    the  DP process reveals two possible solutions: the trivial (Gaussian) fixed point ($u^* = 0$) corresponds to
   a mean-field solution, whereas the non-trivial fixed point, is given by the following approximate coordinate 
\begin{equation}
  u^* = \frac{4 \eps}{3} + 0.92358460(12) \eps^2 - 0.34190(3) \eps^3.
\end{equation}
The Gaussian fixed point is IR attractive for $\eps<0$, which according to Eq.~\eqref{eq:def_eps} corresponds to
higher space dimensions $d>d_c$. On the other hand, the non-trivial  fixed point is IR stable for $\eps>0$, which corresponds
to physically more relevant space dimensions $d<4$.

The existence of IR solution of RG equations implies the scaling behavior of Green functions. From the macroscopic point of view,  
 the parameters of the model can be divided into two groups: IR irrelevant quantities ($D$, $g$, $u$ and $\mu$) and IR relevant (momenta/coordinates,
 frequency/time, $\tau$ and fields).
For a dynamical model \cite{Vasilev2004}, the critical dimension of the IR relevant quantity $Q$ is given by the general formula
\begin{equation}
  \Delta_Q = d_k [Q] + \Delta_\omega d_\omega [Q] + \gamma^*_Q, 
  \label{eq:critic_dims}
\end{equation} 
where the critical dimension of frequency $\Delta_\omega$ is related to the anomalous
dimension $\gamma_D$ by
\begin{equation}
  \Delta_\omega = 2 - \gamma_D^*.
  \label{eq:critic_dim_freq}
\end{equation} 
Normalization conditions are assumed in the form 
\begin{equation}
  \Delta_k = - \Delta_x = 1,
\end{equation}
 and $d_k[Q]$ and $d_{\omega} [Q]$ in Eq.~\eqref{eq:critic_dims} denote canonical dimensions of quantity $Q$ 
(see Tab.~\ref{tab:canon_dim}), $\gamma_Q^*$ is the value of anomalous dimension at the given fixed point.
 The remaining relevant critical dimensions can be expressed through
 critical anomalous dimensions $\gamma_\psi^*$ and $\gamma_\tau^*$
\begin{equation}
   \Delta_{\psi'} = \Delta_{\psi} = \frac{d}{2} + \gamma_\psi^*,\quad
  \Delta_\tau = 2 +  \gamma_\tau^*.
\end{equation}
 Due to a rapidity-reversal symmetry \eqref{eq:symmetry}, only three critical dimensions are thus sufficient to fully describe DP universality class. 

\subsection{Critical Exponents}
 In contrast to multiscaling problems \cite{turbo,Vasilev2004}, such as turbulence, the DP universality class is an
  example of theory exhibiting simple scaling behavior. This amounts to that only finite numbers of 
 independent exponents are necessary to describe unambiguously DP universality class.

 In lower space dimensions $d \leq 2$ theoretical analysis is intricate as infrared singularities are encountered $\eps = 2$ $(d = 2)$, and they can not be dealt by
 RG method~\cite{Janssen2005}. On the
 other hand, for the most realistic three-dimensional DP  process, which corresponds to $\eps = 1$ $(d = 3)$,
 many results from numerical simulations are available nowadays. 
 They exhibit an even higher level of inaccuracy and field-theoretic predictions seem to provide the most reliable estimates. 

In order to describe DP process quantitatively, let us introduce appropriate quantities.
As the most relevant appears the density of active particles $n(t)$ at time $t$ averaged over the entire system, which
can be regarded as the order parameter of DP.
 In active state density, $n(t)$ is expected to scale in an asymptotic limit $t \rightarrow \infty$ according to a power-law
\begin{equation}
  n (\infty) \sim |\tau|^{\beta}.
  \label{eq:order_scaling}
\end{equation}

\begin{table}
\caption{Critical exponents of DP universality class. More details can be found in the literature~\cite{Hinrichsen2000,Janssen2005,Henkel2008,Tauber2014}. }
\begin{ruledtabular}
\begin{tabular}{c | c | c}
Observable & Exponent & Asymptotic relation  \\
\noalign{\smallskip}\hline\noalign{\smallskip}
Order parameter & $\beta = \frac{\Delta_{\psi}}{\Delta_{\tau}}$ & $n(\tau, t \rightarrow \infty) \sim |\tau|^{\beta}$ \\
                & $\alpha = \frac{\Delta_{\psi}}{\Delta_{\omega}}$ & $n(\tau = 0, t ) \sim t^{-\alpha}$ \\
 \noalign{\smallskip}\hline\noalign{\smallskip}
Survival probability & $\beta' = \beta$ & $P(\tau, t \rightarrow \infty) \sim |\tau|^{\beta'}$\\
                & $\delta = \alpha$ & $P(\tau = 0, t) \sim t^{-\delta}$ \\
 \noalign{\smallskip}\hline\noalign{\smallskip}
Mean square radius & $z =\Delta_{\omega}$ & $R^2(\tau =0 ,t) \sim t^{2/z}$ \\
 \noalign{\smallskip}\hline\noalign{\smallskip}
Correlation length & $\nu_{\perp} \equiv \nu  = \frac{1}{\Delta_{\tau}}$ & $\xi(\tau,t \rightarrow \infty) \sim |\tau|^{-\nu_{\perp}}$\\
                  & $\nu_{\parallel}  \equiv z \nu  = \frac{\Delta_{\omega}}{\Delta_{\tau}}$ & $\xi(\tau,t \rightarrow \infty) \sim |\tau|^{-\nu_{\parallel}}$ \\    
 \noalign{\smallskip}\hline\noalign{\smallskip}
Number of active sites & $\theta = \frac{d - 2\Delta_{\psi}}{\Delta_{\omega}}$ & $N(\tau = 0, t) \sim t^{\theta}$ \\
 \noalign{\smallskip}\hline\noalign{\smallskip}
Mean cluster mass & $\gamma = \frac{d + \Delta_{\omega} - 2\Delta_{\psi}}{\Delta_{\tau}}$ & $M(\tau, t \rightarrow \infty) \sim |\tau|^{-\gamma}$ \\
 \noalign{\smallskip}\hline\noalign{\smallskip}
Mean size of cluster & $\sigma = \frac{d+\Delta_{\omega}-\Delta_{\psi}}{\Delta_{\tau}}$ & $S(\tau, t \rightarrow \infty) \sim |\tau|^{-\sigma}$ 
\end{tabular}
\label{tab:index}
\end{ruledtabular}
\end{table}

The scaling behavior of several other quantities can be derived from the two-point Green function. Both from a practical and theoretical points of view
a special role is played by the response
 function $W_{\psi \tpsi} = \langle \psi (t, {\mx}) \tpsi (0, {\bm 0})\rangle$. Its scaling form can take various forms.
 For our purposes here, they are given by
\begin{align}
  W_{\psi \tpsi} (t, \mx, \tau) & = t^{- 2\Delta_{\psi} / \Delta_{\omega}} f \left(  \frac{\mx}{t^{1 / \Delta_{\omega}}}, 
  \frac{\tau}{t^{- \Delta_{\tau /   \Delta_{\omega}}} }\right), 
  \label{eq:asym_w_t} \\ 
  W_{\psi \tpsi} (t, \mx, \tau) & = |\tau|^{ 2 \Delta_{\psi} / \Delta_{\tau}} g \left( \frac{t}{|\tau|^{ -\Delta_{\omega} / \Delta_{\tau} }},
   \frac{\mx}{|\tau|^{- 1 /       \Delta_{\tau}}}\right),
  \label{eq:asym_w_tau}
\end{align}
where we have taken into account the relation $\Delta_{t} = - \Delta_{\omega}$. In DP lattice models~\cite{Henkel2008} the response function
is used in order to define quantities such as the number of active sites $N(t,\tau)$ generated from the origin 
\begin{equation}
  N(t,\tau) = \int \dRM^d x \; W_{\psi \tpsi}(t,{\mx},\tau),
  \label{eq:scale_N}
\end{equation}
which are crucial for seed simulations.

At criticality, the number of active particles displays asymptotic power law behavior governed by the critical exponent $\theta$. 
Substituting Eq.~\eqref{eq:asym_w_t} into
 the definition for $N(t)$ we derive the following scaling form
\begin{equation}
  N(t, \tau) = t^{(d - 2\Delta_{\psi})/\Delta_\omega} F \left( \tau t^{\Delta_{\tau} / \Delta_{\omega}} \right),
  \label{eq:scaling_form_N}
\end{equation}
where the scaling function $F$ is finite at the critical point $\tau = 0$. 
The critical exponent $\theta$ is defined in terms of critical dimensions in the following way
\begin{equation}
  \theta \equiv \frac{d - 2\Delta_{\psi}}{\Delta_\omega} .
  \label{eq:def_theta}
\end{equation}  
We then arrive at a three-loop prediction (the third order in the expansion
parameter $\eps$) 
\begin{equation}
  \theta  = \frac{\eps}{12} + 0.037509726(13) \eps^2 - 0.032978(3) \eps^3.
\end{equation}  
 Another dynamical quantity is the mean square radius $R^2(t)$ of
 spreading particles starting from a single seed at the origin at time $t=0$ 
 \cite{Janssen2005,Henkel2008}.
 It is defined by
\begin{equation}
  R^2(t,\tau) 
  \equiv  \frac{\int \dRM^d x \, \mx^2 W_{\psi \tpsi}(t,\mx, \tau)}{ \int \dRM^d x \;  W_{\psi \tpsi}(t, \mx, \tau)}.
  \label{eq:scale_R}
\end{equation}
In a similar way as has been done for $n(t,\tau)$ we can obtain the asymptotic scaling behavior for $R^2$ in criticality
\begin{equation}
  R^2 \sim t^{2/z}, 
  \label{eq:scale_R_power}
\end{equation}
where the dynamical exponent $z$ in our notation is equal to the critical exponent of frequency, i.e. $ z = \Delta_\omega$.

A two-loop calculation for $z$ can be performed in an analytical fashion~\cite{Janssen1981,Janssen2001,Janssen2005} with the final perturbative prediction
\begin{equation}
   z = 2 - \frac{\eps}{12} - \left(
   \frac{67}{288} + \frac{59}{144}\ln\frac{4}{3}
   \right) \frac{\eps^2}{12}. 	
   \label{eq:z_janssen}
\end{equation}
The corresponding value to the third order in perturbation theory  is determined from the relation \eqref{eq:critic_dims} and reads
\begin{equation}
	z = 2 - \frac{\eps}{12} - 0.02920905(2) \eps^2 
	+ 0.029207(4) \eps^3. 
\end{equation} 
A further quantity, which elucidates the nature of DP universality class, is the mean cluster mass $M$, which is also related to the two-point connected Green function \cite{Henkel2008}
\begin{equation}
  M(\tau) = \int \dRM^d x  \int\limits_{0}^{\infty} \dRM t \; W_{\psi \tpsi} ( \mx, t, \tau).
  \label{eq:cluster_mass}
\end{equation}
Substituting the asymptotic formula \eqref{eq:asym_w_tau} into this
definition and rescaling the time and space variables accordingly leads to
the asymptotic power-law behavior  
\begin{equation}
  M (\tau) 
   \propto |\tau|^{-(d+\Delta_{\omega} -2 \Delta_{\psi})/\Delta_{\tau}},
  \label{eq:cluster_mass_scaling}
\end{equation}
where additional scaling dependence has been suppressed. The definition of the exponent $\gamma$ directly follows
\begin{equation}  
  \gamma = \frac{d+\Delta_{\omega} -2 \Delta_{\psi}}{\Delta_{\tau}} .
  \label{eq:def_gamma}
\end{equation}
From our three-loop calculation,  we obtain the following result
\begin{equation}  
  \gamma = 1 + \frac{\eps}{6}  + 0.06683697(2) \eps^2  
  - 0.036156(4) \eps^3. 
\end{equation}
As has already been mentioned, a complete description of the DP
universality class requires knowledge of exactly three independent critical exponents. Customarily~\cite{Henkel2008,Tauber2014}, the triple
 $(\beta, \nu_{\parallel}, \nu_{\perp})$ is chosen.
 Our calculation shows that to the three-loop approximation they are given by  
\begin{align}
	\beta = 1 &- \frac{\eps}{6} - 0.01128142(2) \eps^2  - 0.015743(3)\eps^3, 
	\label{eq:beta} \\	
	\nu_{\parallel} = 1 &+ \frac{\eps}{12} + 0.02238280(2) \eps^2 - 0.008169(3) \eps^3, 
	\label{eq:nu_parallel} \\
	\nu_{\perp} = \frac{1}{2} &+ \frac{\eps}{16} + 0.021097832(11) \eps^2  - 0.009594(2) \eps^3. 
  \label{eq:nu_perp} 
\end{align}
Additional relations and definitions of critical exponents are summarized in Tab.~\ref{tab:index} and take the following form
\begin{align}
	\alpha = 1 &- \frac{\eps}{4} - 0.012830892(11) \eps^2  - 0.000910(2)\eps^3, 
	\label{eq:other4} \\	
	\sigma =2  &+ \frac{\eps^2}{18}   - 0.051899(8) \eps^3.
	\label{eq:other7}
\end{align}

The critical exponents can be directly compared to previous analytical results~\cite{Janssen1981, Janssen2005} calculated to the two-loop approximation 
(the second order in $\eps$)
\begin{align}
  \beta & = 1 - \frac{\eps}{6} + \left( \frac{11}{12} -\frac{53}{6} \ln \frac{4}{3} \right) \frac{\eps^2}{144},  \\
  \nu_{\parallel} & = 1 + \frac{\eps}{12} + \left(\frac{109}{24} -\frac{55}{12} \ln \frac{4}{3} \right) \frac{\eps^2}{144}, \\
  \nu_{\perp} & = \frac{1}{2} + \frac{\eps}{16} + \left(\frac{107}{32}-\frac{17}{16} \ln \frac{4}{3} \right) \frac{\eps^2}{144}. \end{align}
Their numerical evaluation yields the following expressions
\begin{align}
  \beta & = 1 - \frac{\eps}{6}  -0.0112814234\eps^2,  \\
  \nu_{\parallel} & = 1 + \frac{\eps}{12} +  0.0223828044 \eps^2, \\
  \nu_{\perp} & = \frac{1}{2} + \frac{\eps}{16} + 0.0210978319 \eps^2,
\end{align}
which agrees very well with our results~\eqref{eq:beta}-\eqref{eq:nu_perp}.

In the numerical calculation, there is a non-trivial way to verify the numerical values of critical exponents.
 This is based on a generalized hyperscaling relation \cite{Hinrichsen2000}, which takes the form
\begin{equation}
\theta - \frac{d}{z} = - \frac{\beta+\beta'}{\nu_{\parallel}}.
\end{equation}   
However, this relation is automatically fulfilled once critical exponents are expressed in terms of critical dimensions.

 Another possibility would be to compare numerical results to predictions 
  obtained by Monte-Carlo simulations on a lattice in a certain space dimension \cite{Henkel2008}.

{\subsection{Resummation Results} \label{subsec:resum} }
 It is well-known~\cite{Zinn2002} that  typical perturbative series 
  in quantum (statistical) field
  theory (such as expressions for critical exponents~\eqref{eq:beta}-\eqref{eq:nu_perp}) are asymptotic, rather than convergent.
 There exist powerful mathematical methods suitable for their analysis~\cite{BenderBook}, which might extract useful physical information.
 Therefore, it is reasonable to apply some summation on the part of an asymptotic series. 
In order to attain progress, we apply resummation techniques and compare obtained results.

\begin{table*}
\caption{Evaluation of critical exponents of DP universality class in three space
 dimensions $\eps =1$ $(d = 3)$: Pad\'{e} approximant - $P^N_D$, Pad\'{e} resummation, Direct
 summation, results from Monte Carlo simulations of the contact process
%
%
 and non-perturbative renormalization group. 
%
%
 The symbol $^{*}$ marks value that was calculated using scaling relations~\cite{Hinrichsen2000}.
%
%
 First two rows correspond to Pad\'{e} approximants \cite{Adzhemyan2019,Borinsky2021} calculated from our two-loop
 and three-loop results, respectively.
%
%
  }
\begin{ruledtabular}
\begin{tabular}{c c c c c c c c c}
 & \multicolumn{1}{c}{$\beta = \beta'$} & \multicolumn{1}{c}{$\nu_{\perp}$} & \multicolumn{1}{c}{$\nu_{\parallel}$} & \multicolumn{1}{c}{$\alpha =\delta$} & \multicolumn{1}{c}{$\gamma$} & \multicolumn{1}{c}{$z$} & \multicolumn{1}{c}{$\theta$} & \multicolumn{1}{c}{$\sigma$} \\
\noalign{\smallskip}\hline\noalign{\smallskip}
Pad\'{e} (two-loop) & $0.833(17)$ & $0.580(15)$ & $1.101(15)$ & $0.76(3)$ & $1.23(5)$ & $1.90(2)$ & $0.12(6)$ & $2.04(6)$ \\
\noalign{\smallskip}\hline\noalign{\smallskip}
Pad\'{e} (three-loop) & $0.818(10)$ & $0.581(8)$ & $1.103(6)$ & $0.740(6)$ & $1.23(3)$ & $1.898(14)$ & $0.11(4)$ & $2.03(2)$ \\
\noalign{\smallskip}\hline\noalign{\smallskip}
Ref. \cite{Jensen1992} & $0.813(11)$ & $0.584(6)^{*}$ & $1.11(1)$ & $0.732(4)$ & $1.23(3)^{*}$ & $1.901(5)^{*}$ & $0.114(4)$ & $2.05(4)^{*}$ \\
\noalign{\smallskip}\hline\noalign{\smallskip}
Ref. \cite{Vojta2012} & $0.815(2)$ & $0.5826(9)$ & $1.106(2)$ & $0.7367(6)$ & $1.224(5)^{*}$  & $1.8986(8)$ & $0.1062(4)$ & $2.039(4)^{*}$ \\
\noalign{\smallskip}\hline\noalign{\smallskip}
Ref. \cite{Dickman1999} & $0.809(3)^{*}$ & $0.580(2)^{*}$ & $1.114(4)$ & $0.7263(11)$ & $1.237(6)^{*}$ & $1.919(4)^{*}$ & $0.110(1)$ & $2.046(8)^{*}$ \\
\noalign{\smallskip}\hline\noalign{\smallskip}
Ref. \cite{Deng2013} & $0.818(4)$ & $0.582(2)$ & $1.106(3)$ & $0.7398(10)$ & $1.216(10)^{*}$  & $1.8990(4)$ & $0.1057(3)$ &  $2.034(8)^{*}$ \\
\noalign{\smallskip}\hline\noalign{\smallskip}
Ref. \cite{canet2004} & $0.782$ & $0.548$ & $1.046^{*}$  & $0.747^{*}$ & $1.126^{*}$ & $1.909$ & $0.0764^*$  & $1.908^{*}$ \\
\end{tabular}
\label{tab:exp_results}
\end{ruledtabular}
\end{table*}

 The final results of critical exponents are calculated as an asymptotic series in formally small $\eps = 4 - d$.  
To obtain final values in the physically relevant space dimension $d=3$ ($\eps = 1$) we use Pad\'{e} approximants with the strategy described in \cite{Adzhemyan2019,Borinsky2021}.
These values of critical exponents  can be found in Table~\ref{tab:exp_results}, where they are compared with the results from Monte Carlo simulation for the contact process \cite{Jensen1992,Vojta2012} and
with results obtained from the high-precision Monte-Carlo analysis~\cite{Dickman1999,Deng2013}.
The latter belongs to the DP universality class, and calculated critical exponents are in  agreement within  corresponding error bars.

{\section{Conclusion}	\label{sec:conclusion} }
In this paper, we have calculated the critical exponents of DP universality class up to the third order of perturbation theory. 
We have formulated DP process within the field-theoretic framework and performed a full UV renormalization analysis. We have limited
ourselves to compute only three critical exponents needed for a description of the DP process. 
In particular, we have found perturbative $\eps$-expansion for three independent critical
 exponents $\beta, \nu_{\perp}, \nu_{\parallel}$ (or $\theta, z, \gamma$) to the
third order in $\eps$. Our results are in agreement with the two-loop analytic calculations.

 In calculation of the RG constants, we have identified three distinguished properties, which has helped us simplify 
difficult technical part related to analytical treatment of three-loop  diagrams.
 Moreover, we have used them for independent crosschecks to verify obtained results.
  In the space dimension $d = 3$ ($\eps = 1$), we have carried out
 the resummation technique using Pad\'{e} approximants. Obtained results are in accordance with the prediction of critical exponents from a Monte-Carlo
  simulations. We expect that in the next order of perturbation theory, the evaluation of critical exponents may
   provide even better results than simulations. In addition, in higher orders, it is possible to use other resummation methods built on Pad\'{e} approximants 
   with the Borel-Leroy transformation \cite{Adzhemyan2019} or the Conformal mapping \cite{Zinn2002}. 

This study presents a step towards enhancing multi-loop calculations in non-equilibrium physics. It could help to solve other more
involved dynamical models. There are several other directions in which this work can be extended. 
 One conceivable task would be an analysis of universal amplitude ratios. Another potential direction would be even more sophisticated
 four-loop calculation, which would be beneficial for asymptotic analysis. 

\begin{acknowledgments}
The work was supported by VEGA grant No. 1/0535/21 of the Ministry of Education, Science, 
Research and Sport of the Slovak Republic and by the grant of the Slovak
Research and Development Agency under the contract No. APVV-21-0319.
\end{acknowledgments}
\appendix
{\section{Symmetry properties}	\label{sec:appendixa} }

In this Appendix, the relationships between self-energy and vertex diagrams will be obtained, which were used to verify the calculations. 

\begin{figure}
\includegraphics[width=8cm]{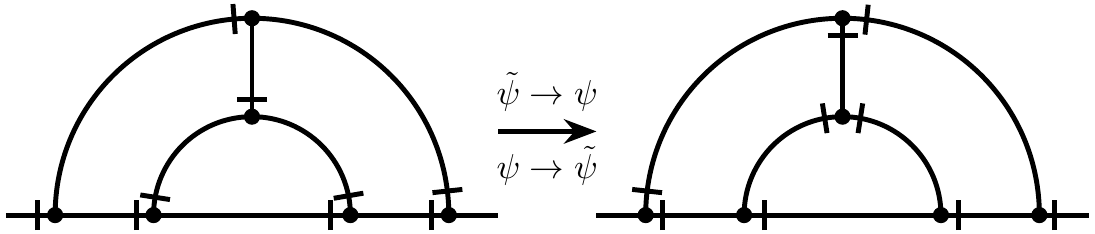}
\caption{Three-loop diagrams of DP process for two-point Green function $\Gamma^{(1,1)}$
bounded by rapidity-reversal symmetry \eqref{eq:symmetry}.}
\label{fig:time_sym} 
\end{figure}

The symmetry relations \eqref{eq:symmetry} with respect to time reversal can be used to find connections between diagrams, e.g., it is possible to find identical diagrams among self-energy diagrams and vertex diagrams. For self-energy diagrams, this follows directly from the fact that transformation \eqref{eq:symmetry} transforms the self-energy graphs 
$\langle \psi \tpsi \rangle_{1PI}$
into
$\langle \psi \tpsi \rangle_{1PI}$.
In Fig. \ref{fig:time_sym}, the example of such a transformation is demonstrated, explicitly showing the equality of the depicted diagrams.

In some cases, such a transformation does not provide information - it leads to identical equality, as, for example, for the one-loop diagram \eqref{eq:self_energy1} and both two-loop diagrams \eqref{eq:self_energy2}. For three-loop self-energy functions, in addition to the relations  determined in  Fig. \ref{fig:time_sym}, there are also $4$ 
non-trivial identities.
    
\begin{figure}
\includegraphics[width=8cm]{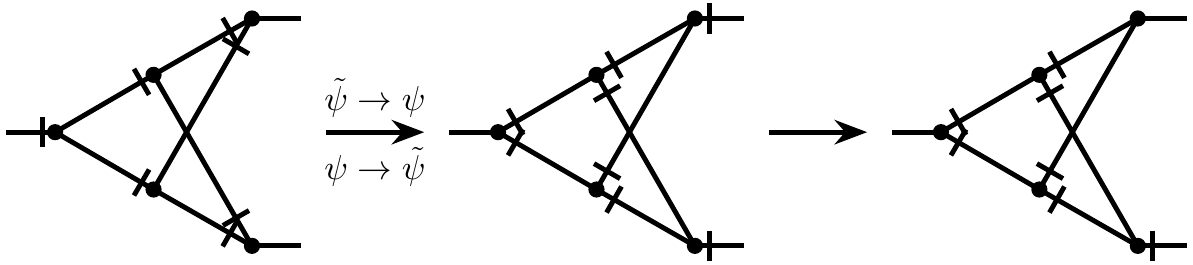}
\caption{Example of a diagram of
$\Gamma^{(1,2)}$,
which after the indicated substitution can not be represented as a diagram of  
$\Gamma^{(1,2)}$ and has to be discarded from a consequent analysis.
}
\label{fig:time_sym3} 
\end{figure}

By means of relations \eqref{eq:symmetry} the vertex function 
$\Gamma^{(1,2)}$ is transformed to the 
$\Gamma^{(2,1)}$. 
\begin{figure}
\includegraphics[width=8cm]{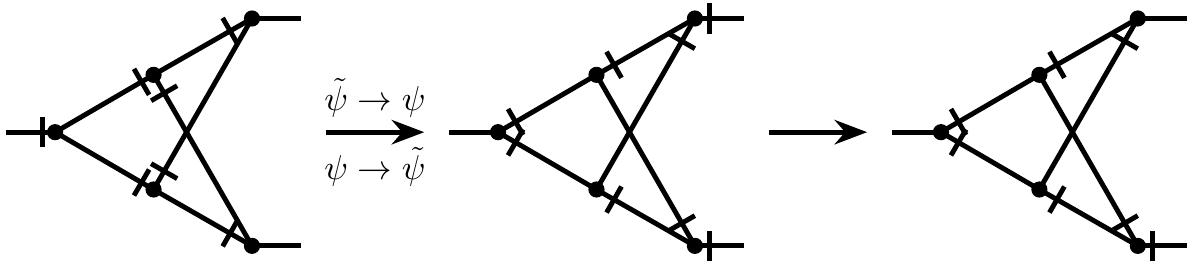}
\caption{Example of a diagram for Green function 
$\Gamma^{(1,2)}$,
which after substitution can be represented as another existing diagram of
$\Gamma^{(1,2)}$. }
\label{fig:time_sym2} 
\end{figure}
Fig. \ref{fig:time_sym2}  shows how it is possible to replace the upper outer tail $\tpsi$ with $\psi$, which gives information about the equality of two different diagrams  
$\Gamma^{(1,2)}.$ 
This procedure is not informative for every  vertex diagram, for example, in Fig. \ref{fig:time_sym3}
it is impossible to replace any of the outer tails $\tpsi$ with $\psi$ in the second diagram, since this would give rise to a triple vertex 
$\Gamma^{(0,3)}$
that is, however, absent in theory. This situation takes place for all diagrams
$\Gamma^{(1,2)}$
in which both outer tails are connected to two fields $\tpsi$, as in the example considered. As a result, for the 1PI diagrams of the Green function
$\Gamma^{(1,2)}$, 
$52$ 
equalities of identical pairs of diagrams were obtained.

Some additional connections between the values of three-loop diagrams can be obtained using the equality 
\begin{equation}
  \includegraphics[width=4cm]{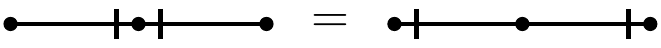}.
  \label{eq:special}
\end{equation}
One of its consequences is the equality of three pairs 1PI diagrams
\begin{align}
\raisebox{-6.4ex}{\includegraphics[width=3.2cm]{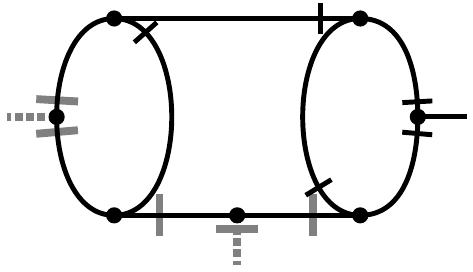}}  =  
\raisebox{-6.4ex}{\includegraphics[width=3.2cm]{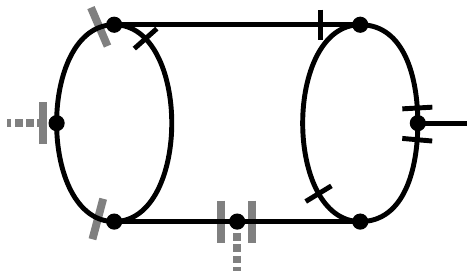}} \ ,
\\
\raisebox{-10ex}{\includegraphics[width=3.2cm]{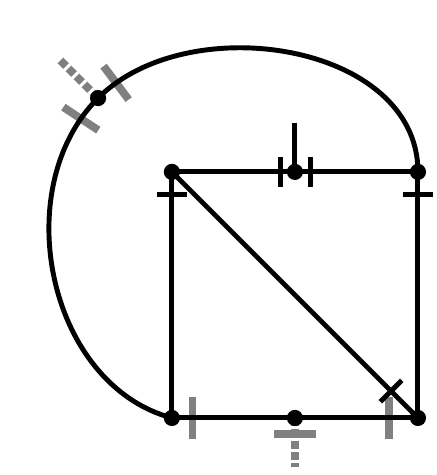}} \ =
\raisebox{-10ex}{\includegraphics[width=3.2cm]{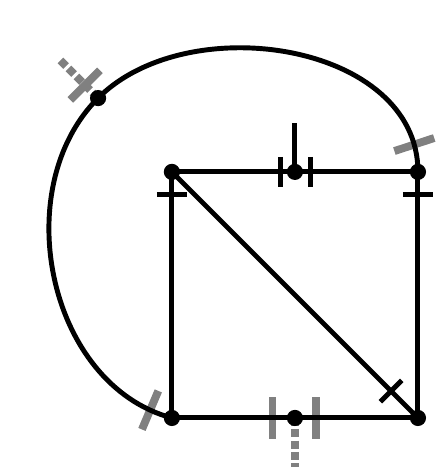}} \ ,
\\
\raisebox{-10ex}{\includegraphics[width=3.2cm]{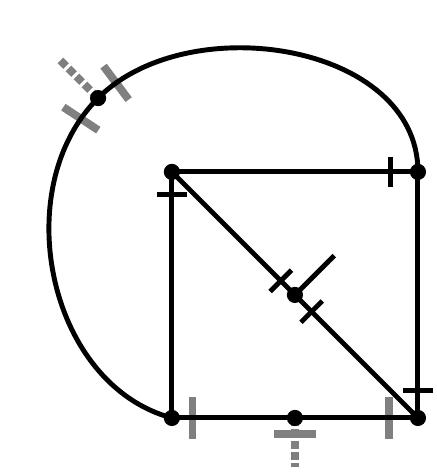}} \ =
\raisebox{-10ex}{\includegraphics[width=3.2cm]{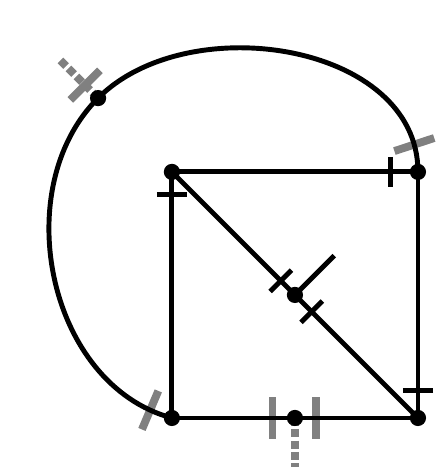}} \ ,
\end{align} 
where the dotted external lines correspond to the use of equality \eqref{eq:special}. Response fields should be added appropriately 
using the graphical rule for the interaction vertex (see Fig. \ref{fig:DP_rules}).

A feature of the  model \eqref{eq:action_0} is the presence of a single propagator \eqref{eq:prop_omega}. This allows us to reconstruct all diagrams of 
1PI functions
\eqref{eq:Gamma1} and \eqref{eq:Gamma3} from the diagrams of function \eqref{eq:Gamma4}. Differentiating each propagator in the diagrams of the 1PI function \eqref{eq:Gamma1} with respect to the external frequency, through which the frequency flows, and taking into account \eqref{eq:prop_omega}, gives
\begin{equation} 
\partial_{i\omega} \left[ \frac{1}{-i(\omega+\omega') + \eps_k}\right] =  \left[\frac{1}{-i(\omega+\omega') + \eps_k} \right]^2,
\label{eq:dif_omega}
\end{equation}
where $\omega'$ and $k$ are the frequency and momentum of integration and
 $\eps_k$ was introduced in Eq.~\eqref{eq:prop_omega}.
 In diagrammatic language, this relation can be written as
\begin{equation}
\raisebox{-1.ex}{\includegraphics[width=6cm]{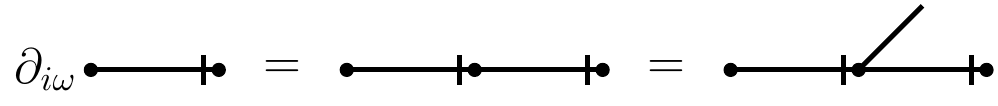}} .
\label{eq:dif_omega_diag}
\end{equation}

The first equality is directly equivalent to relation \eqref{eq:dif_omega}, the additional line in the second equality depicts the outer tail of the diagram, its insertion does not change the integrand for the diagram and turns it into the corresponding diagram for the three-point function.
Note that when calculating function \eqref{eq:Gamma1}, the external frequency after differentiation $\partial_{i\omega}$ is assumed to be equal to zero, as well as when calculating function \eqref{eq:Gamma4}.

\begin{figure}
\includegraphics[width=5cm]{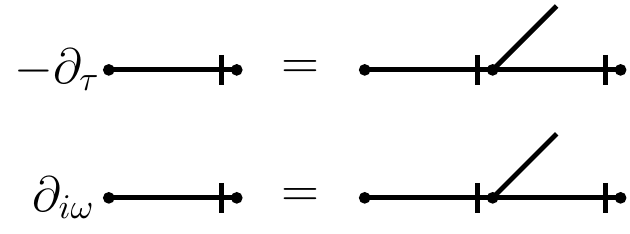}
\caption{Graphical representation of differentiation of DP propagator with respect to the $\tau$ (top), and
   with respect to the frequency $\omega$ (bottom).}
\label{fig:derivative} 
\end{figure}

\begin{figure}
\includegraphics[width=8.5cm]{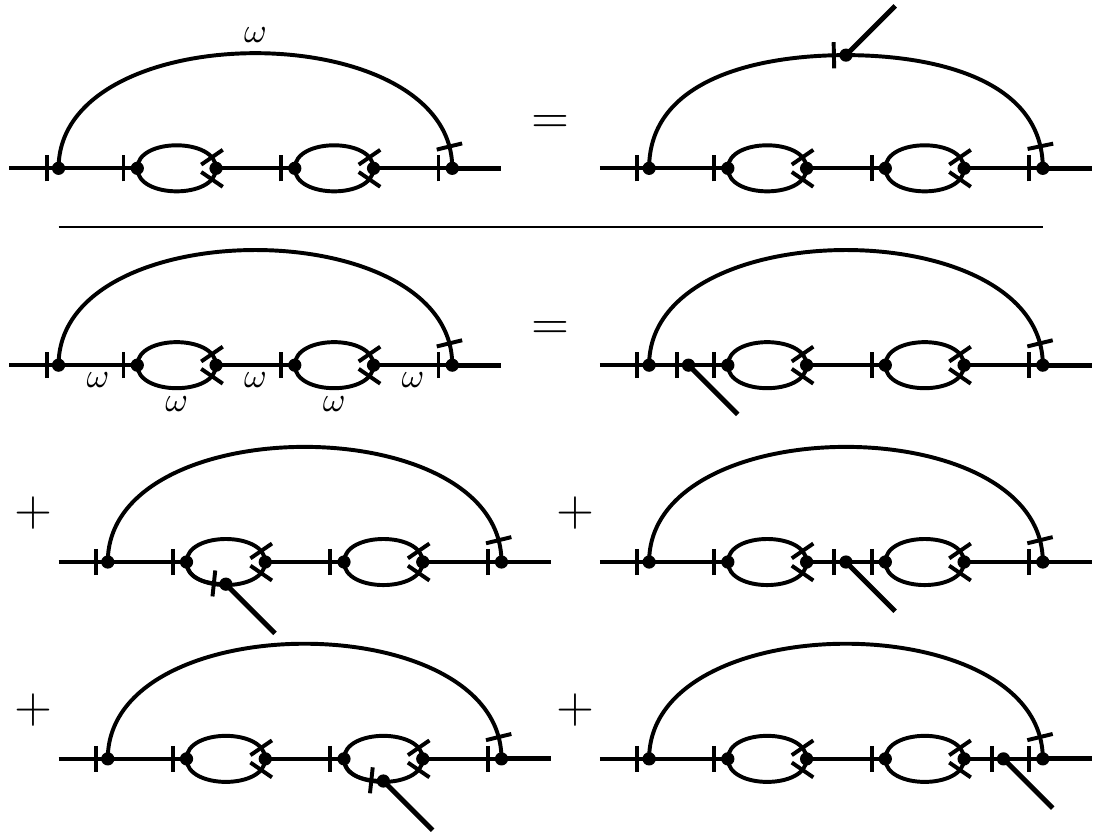}
\caption{Specific two-point diagram of DP process with two different possibilities
for the flow of the external frequency 
$\omega$. Propagators through which $\omega$ is flowing are subsequently replaced in all possible manner to yield
a sum of corresponding three-point diagrams.}
\label{fig:frequency} 
\end{figure}

Thus, differentiation with respect to frequency when calculating the diagrams of function \eqref{eq:Gamma1} can be replaced by the sum of diagrams of the form \eqref{eq:Gamma4}. Fig.~\ref{fig:frequency} displays an example of such a replacement for two different directions of external momentum  leakage. Since the result does not depend on this direction, the right-hand sides of the equalities have to coincide, which can serve as a check of the correctness of the calculation of three-point diagrams.

Equalities \eqref{eq:dif_omega}, \eqref{eq:dif_omega_diag} remain valid after the replacement $\partial_{i\omega} \rightarrow  \partial_{\tau_0}/ D_0$.  This means that the sum of the contributions when differentiating each line when calculating the
1PI function \eqref{eq:Gamma3} can be replaced by the sum of the corresponding three-tailed diagrams in function \eqref{eq:Gamma4}.

\bibliographystyle{apsrev}
\bibliography{mybib}

\end{document}